\documentclass[longauth]{aa}
\pdfoutput=1

\usepackage{graphicx}
\usepackage{subfig}
\usepackage[varg]{txfonts}
\usepackage{lscape}
\usepackage{color}
\usepackage{longtable}

\def\m2s2{\hbox{\,m$^{2}$\,s$^{-2}$}} %m2.s -2
\def\fSNCJ{f(\mathrm{SN} | \mathrm{CJ})}
\def\fSN{f(\mathrm{SN})}
\def\fCJ{f(\mathrm{CJ})}
\def\fCJSN{f(\mathrm{CJ} | \mathrm{SN})}

\begin{document}

   \title{The GAPS programme at TNG\thanks{Based on: observations made with the Italian \textit{Telescopio Nazionale Galileo} (TNG), operated on the island of La Palma by the INAF - \textit{Fundaci\'on Galileo Galilei} at the \textit{Roque de Los Muchachos} Observatory of the \textit{Instituto de Astrof\'isica de Canarias} (IAC).}}
   \subtitle{XLVI. Deep search for low-mass planets in late-dwarf systems hosting cold Jupiters}
   \titlerunning{Late-type Jupiters}

   \author{M. Pinamonti\inst{\ref{inst1}}
          \and
          D. Barbato\inst{\ref{inst2}}
          \and
          A. Sozzetti\inst{\ref{inst1}}
          \and
%           A. S. Bonomo\inst{\ref{inst1}}
%           \and
          L. Affer\inst{\ref{inst3}}
          \and
          S. Benatti\inst{\ref{inst3}}
          \and
          K. Biazzo\inst{\ref{inst4}}
          \and
          A. Bignamini\inst{\ref{inst5}}
          \and
          F. Borsa\inst{\ref{inst6}}
          \and
          M. Damasso\inst{\ref{inst1}}
          \and
          S. Desidera\inst{\ref{inst7}}
          \and
          A. F. Lanza\inst{\ref{inst4}}
          \and
          J. Maldonado\inst{\ref{inst3}}
          \and
          L. Mancini\inst{\ref{inst1},\ref{inst8},\ref{inst9}}
          \and
          L. Naponiello\inst{\ref{inst1},\ref{inst8},\ref{inst10}}
          \and
          D. Nardiello{\ref{inst7}}
          \and
          M. Rainer\inst{\ref{inst6}}
          \and
          L. Cabona\inst{\ref{inst7}}
          \and
          C. Knapic\inst{\ref{inst5}} 
          \and
          G. Andreuzzi\inst{\ref{inst11},\ref{inst15}}  
          \and
          R. Cosentino\inst{\ref{inst11}}   
          \and
          A. Fiorenzano\inst{\ref{inst11}}           
          \and
          A. Ghedina\inst{\ref{inst11}}  
          \and
          A. Harutyunyan\inst{\ref{inst11}}        
          \and
          V. Lorenzi\inst{\ref{inst11},\ref{inst16}} 
          \and
          M. Pedani\inst{\ref{inst11}}     
          \and
          R. Claudi\inst{\ref{inst7}}     
          \and
          E. Covino\inst{\ref{inst12}}   
          \and
          A. Maggio\inst{\ref{inst3}}  
          \and
          G. Micela\inst{\ref{inst3}} 
          \and
          E. Molinari\inst{\ref{inst13}} 
          \and
          I. Pagano\inst{\ref{inst4}}
          \and
          G. Piotto\inst{\ref{inst14}}
          \and
          E. Poretti\inst{\ref{inst6},\ref{inst11}}
          }

   \institute{INAF - Osservatorio Astrofisico di Torino, Via Osservatorio 20, I-10025 Pino Torinese, Italy\\
              \email{m.pinamonti.astro@gmail.com}\label{inst1}
         \and
             Observatoire de Genève, Université de Genève, 51 Chemin des Maillettes, CH-1290 Sauverny, Switzerland\label{inst2}
         \and
             INAF - Osservatorio Astronomico di Palermo, piazza del Parlamento 1, I-90134 Palermo, Italy\label{inst3}
         \and
             INAF - Osservatorio Astrofisico di Catania, Via S. Sofia 78, I-95123 Catania, Italy\label{inst4}
         \and
             INAF - Osservatorio Astronomico di Trieste, via G. B. Tiepolo 11, I-34143 Trieste, Italy\label{inst5}
         \and
             INAF - Osservatorio Astronomico di Brera, Via E. Bianchi 46, I-23807 Merate, Italy\label{inst6}
         \and
             INAF - Osservatorio Astronomico di Padova, vicolo dell'Osservatorio 5, I-35122 Padova, Italy\label{inst7}
         \and
             Dipartimento di Fisica, Universit\`a di Roma ``Tor Vergata'', Via della Ricerca Scientifica 1, I-00133 Roma, Italy\label{inst8}
         \and
             Max Planck Institute for Astronomy, Kõnigstuhl 17, DE-69117 Heidelberg, Germany\label{inst9}
         \and
             Dipartimento di Fisica e Astronomia, Universit\`a di Firenze, Largo Enrico Fermi 5, I-50125 Firenze, Italy\label{inst10}
         \and
             Fundaci\'on Galileo Galilei - INAF, Ramble Jos\'e Ana Fernandez P\'erez 7, E-38712 Bre\~na Baja, TF, Spain\label{inst11}
         \and
             INAF - Osservatorio Astronomico di Roma, Via Frascati 33, I-00078 Monte Porzio Catone, Italy\label{inst15}
         \and
             Instituto de Astrof\,isica de Canarias (IAC), E-38205 La Laguna, Tenerife, Spain\label{inst16}
         \and
             INAF - Osservatorio Astronomico di Capodimonte, Salita Moiariello 16, I80131, Napoli, Italy\label{inst12}
         \and
             INAF - Osservatorio di Cagliari, via della Scienza 5, I09047 Selargius, Italy\label{inst13}
         \and
             Dipartimento di Fisica e Astronomia, Universit\`a di Padova, via Marzolo 8, I-35131 Padova, Italy\label{inst14}
         }

   \date{Received <date> /
      Accepted <date>}

  \abstract
  % context heading (optional)
  % {} leave it empty if necessary  
   {With the growth of comparative exoplanetology, it is increasingly clear that the relationship between inner and outer planets plays a key role in unveiling the mechanisms governing formation and evolution models. For this reason, it is important to probe the inner region of systems hosting long-period giants, in search of undetected lower-mass planetary companions.
   }
  % aims heading (mandatory)
   {We present the results of a high-cadence and high-precision radial velocity (RV) monitoring of 3 late-type dwarf stars hosting long-period giants with well-measured orbits, in order to search for short-period sub-Neptunes (SN, $M \sin i < 30$ M$_\oplus$).
   }
  % methods heading (mandatory)
   {Building on the results and expertise of our previous studies, we carry out combined fits of our HARPS-N data with literature RVs. We use MCMC analyses to refine the literature orbital solutions and search for additional inner planets, applying Gaussian Process regression techniques to deal with the stellar activity signals where required. We then use the results of our survey to estimate the frequency of sub-Neptunes in systems hosting cold-Jupiters, $\fSNCJ$, and compare it with the frequency around field M-dwarfs, $\fSN$.}
  % results heading (mandatory)
   {We identify  a new short-period low-mass planet orbiting GJ 328, GJ 328\,c, with $P_c = 241.8^{+1.3}_{-1.7}$ d and $M_c \sin i = 21.4^{+ 3.4}_{- 3.2}$ M$_\oplus$. We moreover identify and model the chromospheric activity signals and rotation periods of GJ 649 and GJ 849, around which no additional planet is found. Then, taking into account also planetary system around the previosuly-analyzed low-mass star BD-11 4672, we derive an estimate of the frequencies of inner planets in such systems. In particular $\fSNCJ = 0.25^{+0.58}_{-0.07}$ for mini-Neptunes ($10$ M$_\oplus < M \sin i < 30$ M$_\oplus$, $P < 150$ d), marginally larger than $\fSN$. For lower-mass planets ($M \sin i < 10$ M$_\oplus$) instead $\fSNCJ <0.69$, compatible with $\fSN$.}
  % conclusions heading (optional), leave it empty if necessary 
   {In light of the newly detected mini-Neptune, we find tentative evidence of a positive correlation between the presence of those planets and that of inner low-mass planets, $\fSNCJ > \fSN$. This might indicate that cold Jupiters have an opposite influence in the formation of inner sub-Neptunes around late-type dwarfs as opposed to their solar-type counterparts, boosting the formation of mini-Neptunes instead of impeding it.
   }
   
   \keywords{techniques: radial velocities - stars: individual: GJ 328 - stars: individual: GJ 649 - stars: individual: GJ 849 - stars: activity - planets and satellites: detection}

   \maketitle
%
%-------------------------------------------------------------------

\section{Introduction}

% \hrule

The growing abundance of known exoplanetary systems has fostered an increasing number of comparative studies on their mass distribution and global architecture \citep{winnfab15,hobsongomez2017}, many of which focus on the role of multiplicity and architecture in supporting or disproving the current and competing formation models for both giant and terrestrial planets \citep[e.g.][]{raymondetal2008,cossouetal2014,schlaufman2014,morbidelliraymond2016}. A key parameter in discriminating between different formation models is the fraction of planetary systems featuring both gas giants and lower-mass planets.
A still-debated aspect of planetary formation models is that of the formation of sub-Neptunes in systems hosting long-period giant planets\footnote{Hereafter we will adopt a broad definition of ``sub-Neptune'' commonly used in the literature, i.e. $M \sin i < 30 $ M$_\oplus$, that encompasses planets commonly referred to as mini-Neptunes ($10$ M$_\oplus < M \sin i < 30$ M$_\oplus$) and super-Earths ($M \sin i < 10 $ M$_\oplus$). Moreover, we will define long-period giant planets, or cold Jupiters, as planets with semi-major axis $a>1$ AU, and $M \sin i > 0.1 $ M$_\text{J}$.}: if low-mass planets embryos form in situ, i.e., close to their host star, the presence of an outer giant planet could block the inward flux of pebbles, impeding a significant growth of the embryo above $\simeq 1$  M$_\oplus$, \citep{lambrechtsetal2019}; this would cause an anti-correlation between the presence of outer giant planets and inner super-Earths, with the important difference that the presence of terrestrial or sub-terrestrial planets would not be significantly affected. On the other hand, other models of in-situ formation of super-Earths require massive protoplanetary disks \citep{chianglaughlin2013}, which would consequently produce a correlated population of  giant planets.
Instead, in the inward migration model, planetary embryos form in the outer protoplanetary disk and migrate inwards to become hot mini-Neptunes: however, if the innermost embryo grows into a gas-giant planet before starting to migrate, it will block the inward migration of more distant cores, creating a dynamical barrier \citep{izidoroetal2015}, and causing an anti-correlation between the gas giant planets and inner low-mass planets. However, if cold Jupiters originate instead from the more distant cores, the cores formed near the snowline can freely migrate inward \citep{bitschetal2015}. Even in the presence of a giant planet acting as migration barrier, simulations predict that a small fraction of planetary cores could be able to ``jump'' the obstacle \citep{izidoroetal2015}, producing a small population of close-in mini-Neptunes in systems with Jupiter-like planets.

To investigate the occurrence of Solar-Type stars hosting inner sub-Neptunes in the presence of outer giant planets, \citet{barbatoetal2018} previously RV monitored 20 bright stars orbited by at least one long-period giant planet with low-to-moderate eccentricity with HARPS at ESO-3.6m, in order to search for inner additional lower-mass planets, and produce a first assessment of their frequency around Solar-type stars. However, no convincing evidence of additional inner low-mass planets in the selected systems was found, most probably due to the low number of data collected, limiting the sample completeness for planets below $M sin i < 30$ M$_\oplus$. This stresses the importance of collecting a high number of dense observations in order to detect any possible low-mass planet present in the sample.
Moreover, the importance of continuing observing at high precision and cadence planet-hosting stars that were already the subject of planet-searching radial velocity surveys has been clearly showcased by the TESS space telescope and its recent discovery of transiting hot super-Earths around stars hosting long-period giant planets \citep[e.g.][]{gandolfietal2018,huangetal2018,lilloboxetal2023}.

Different previous studies reported a high frequency of cold-Jupiters companions in transiting sub-Neptune systems \citep{zhuwu2018,bryanetal2019}. However, since the observable directly related to the formation models is in fact the frequency of sub-Neptunes in cold-Jupiter systems, $\fSNCJ$, these studies might be painting a blurry picture: it is not straightforward to deduce $\fSNCJ$ from the opposite measurement  $\fCJSN$, i.e. the frequency of cold Jupiters in sub-Neptune systems, since its calculation depends on the respective absolute frequencies of both classes of planets, $\fCJ$ and $\fSN$. Both \citet{zhuwu2018} and \citet{bryanetal2019}, assuming $\fCJ$ and $\fSN$ from other surveys, obtain $\fSNCJ  \simeq 100\%$. However, adopting absolute frequencies from other surveys could introduce bias in the calculations, since different samples of stars might not be affected by the same formation and evolution, and thus host different planetary populations with different architectural properties.
Even with high-precision RV time series, studies of the frequency of cold Jupiters in systems with known low-mass planets can be biased due to the short temporal baseline of the data. In short time series, long-period companions can only be identified from RV linear trends, which could affect the results due to the inherent degeneracies in such detections: e.g. the derived orbital parameters for cold Jupiters identified from the RV trends by \citet{bryanetal2019}  have very large uncertainties, spanning on average 2 orders of magnitude in mass and 1 order of magnitude in semi-major axis.

On the other hand, an estimate of $\fSNCJ$ could be computed from systems with known long-period giant planets with well-resolved orbits, and would not suffer from the same biases discussed before. Moreover, the measured frequency would be directly comparable with theoretical predictions \citep[e.g.][]{izidoroetal2015}, without requiring to resort to inferential statistics. However, systematic studies in this direction have been rare in the recent literature \citep{barbatoetal2018,rosenthaletal2022}.

The observational programme Global Architecture of Planetary Systems \citep[GAPS, see][]{covinoetal2013,desideraetal2013} aims at investigating the variety and origins of the architecture of exoplanetary systems with the High Accuracy Radial velocity Planet Searcher in the Northern hemisphere \citep[HARPS-N,][]{cosentinoetal2012} at the Telescopio Nazionale Galileo (TNG) in La Palma. Within this context, a subset of planet-hosting late-K and M dwarfs has been selected and observed in order to search for inner super-Earth and mini-Neptune planetary companions to outer long-period giants and assessing the impact of inefficient migration of planets formed beyond the snowline of such dwarf stellar hosts. The size of the sample was limited by the low frequency of giant planets around late-type stars, with only four target accessible in the Northern hemisphere.

In this paper, we present the analyses of three M dwarfs hosting long-period giant planets, in search for inner low-mass planets. This work follows the analysis carried out in \citet{barbatoetal2020} for the low-mass star BD-11 4672, which is the fourth target observed within the survey. Moreover, we present a statistical analysis of the detection limits of the observed sample, in order to estimate the frequency $\fSNCJ$ around late-type dwarfs, following the Bayesian approach adopted in \citet{pinamontietal2022}. 

\begin{figure}
   \centering
   \includegraphics[width=9cm]{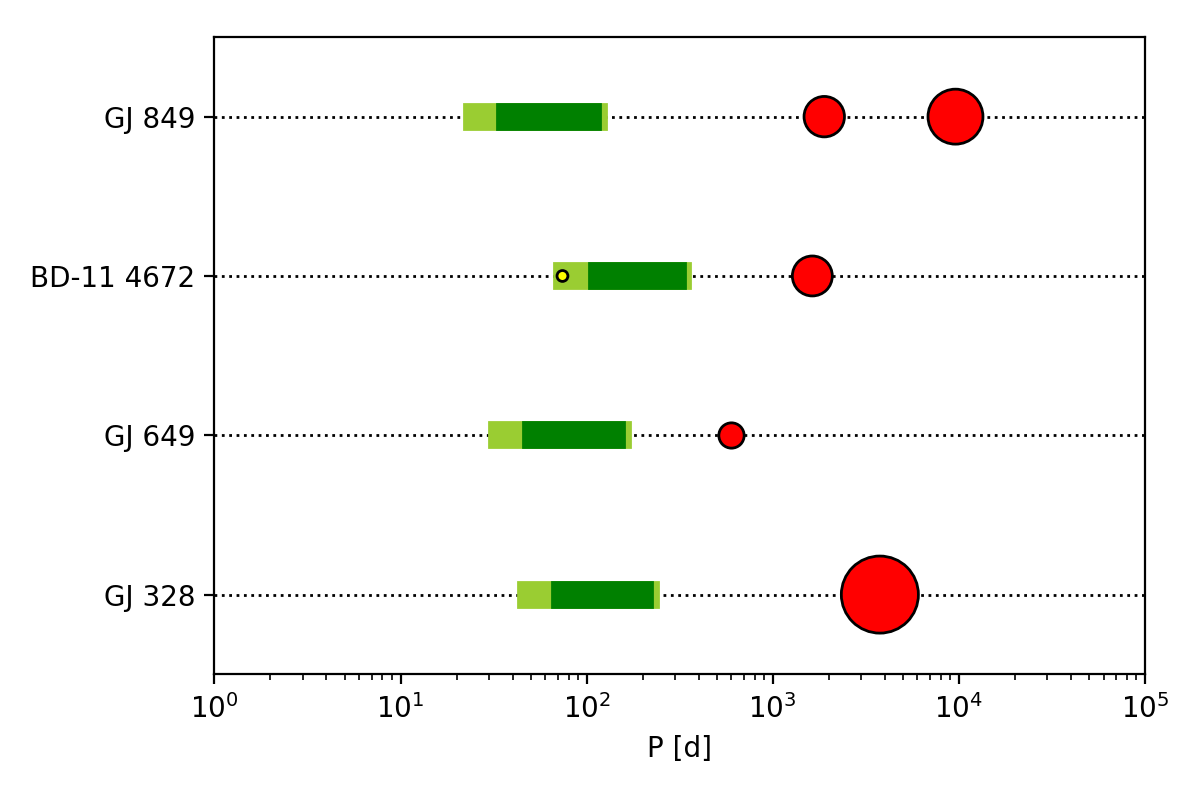}
      \caption{Orbital architecture of the known planets in the observed systems. The giant planets in the sample are shown as red circles, with the symbol size proportional to the minimum planetary mass. The yellow circle show the Neptune-mass planet BD-114672\,c \citep{barbatoetal2020}. The conservative and optimistic limits of the habitable zone of each system \citep{kopparapuetal2013} are shown as thick dark green and light green bands, respectively.}
         \label{fig:systems_lit}
\end{figure}

GJ 328 is a nearby ($d = 20$ pc) early-M dwarf, of type M0. \citet{robertsonetal2013b} detected a long-period ($P = 11$ yr) 2 Jupiter-mass planet on an eccentric orbit, namely GJ 328\,b. At the time of publication, it was the most massive most distant planet found orbiting a red dwarf star. \citet{robertsonetal2013b} studied the magnetic activity of the star and argued that it produced a significant RV signal with an amplitude of $6$-$10$ m s $^{-1}$, which was partly masquerading the eccentricity of GJ 328\,b's orbit. GJ 649 is a nearby ($d = 10$ pc) M1 dwarf, around which \citet{johnsonetal2010} detected a Saturn-mass planet ($m_b \sin i = 0.328$ M$_\text{Jup}$) on an eccentric ($e=0.30$) 600 d orbit. \citet{johnsonetal2010} find also evidence of stellar rotation ($P_\text{rot} = 24.8 \pm 1.0$ d) and long-term magnetic evolution from spectroscopic and photometric monitoring of the star. Finally, GJ 849 is a nearby ($d = 8.8$ pc) M3.5 dwarf, around which one of the first M-dwarf planets was discovered \citep{butleretal2006}: $m_b \sin i = 0.82$ M$_\text{Jup}$, $P_b = 1890 \pm 130$ d. A linear trend was observed in the RV data by \citet{butleretal2006}, which was later confirmed to be due to an outer planetary companion by \citet{fengetal2015}, with minimum mass $m_c \sin i = 0.944 \pm 0.07$ M$_\text{Jup}$ and period $P_c = 15.1 \pm 0.66$ yr.
The planets already known from the literature orbiting the targets are shown in Fig. \ref{fig:systems_lit}.

In Sect. \ref{sec:time_series} we describe the Doppler measurements of our targets collected for this analysis.
% The independent analyses of the two photometric datasets are presented in Sect. \ref{sec:photo_analysis}.
We then describe our analyses of the RV data and stellar activity indices of the targets in Sect. \ref{sec:spectr_analyses}, following with the description of the adopted technique to derive the detection limits and planetary occurrence rates in Sect. \ref{sec:detection_occurrence}, where we also discuss our estimates of the occurrence rates in comparison with the literature.
Finally, we summarize our findings in Sect. \ref{sec:conclusions}.

%--------------------------------------------------------------------

\section{Spectroscopic observations}
\label{sec:time_series}

As part of the GAPS RV programme, the target stars have been observed from  April 2018 to May 2021, with the HARPS-N spectrograph, connected by fibers to the Nasmyth B focus through a Front End Unit of the 3.58m Telescopio Nazionale Galileo (TNG) in La Palma, Spain. HARPS-N is a fiber-fed, cross-dispersed echelle spectrograph with a spectral resolution of $115\;000$, covering a wavelength range from 3830 to 6900 $\AA$ \citep{cosentinoetal2012}. Observations were performed using a fixed integration time of 900 s to obtain data of sufficient per-pixel signal-to-noise ratio (average SNR $> 50$ at 550 nm) and to average out potential short-term periodic oscillations of the star \citep{dumusqueetal2011}.
As discussed in previous studies of M dwarfs with HARPS-N within the HADES program \citep[e.g.][]{afferetal2016}, the spectra were collected without simultaneous Th-Ar calibration, that could contaminate the CaII H$\&$K lines, which are an important activity indicator in M dwarfs \citep{forveilleetal2009,lovisetal2011}. Furthermore, previous analyses proved that no significant instrumental drift is observed even without simultaneous calibration \citep{pergeretal2017,afferetal2019,pinamontietal2019}.
The RVs collected within our survey are combined in our analysis with archival public data from different instruments, including those used in the original discovery of the cold Jupiters present in the observed systems. In the following, we discuss in some detail the spectroscopic time series collected for each analyzedsystem.

Whenever available, we used data from the following instruments: the High Resolution Spectrograph \citep[HRS,][]{tull1998} on the 9.2 m HET \citep{ramseyetal1998} at McDonald
Observatory; the Robert J. Tull Coudé spectrograph \citep{tulletal1995} on McDonald's 2.7 m Harlan J. Smith Telescope (HJST); the High Resolution Echelle Spectrometer \citep[HIRES,][]{ vogtetal1994} spectrograph on the 10 m Keck I telescope at Mauna Kea; the Spectrographe pour l'Observation des Ph\'enom\`enes des Int\'erieurs stellaires et des Exoplan\`etes \citep[SOPHIE,][]{bouchyetal2013} at the 1.93m OHP telescope; the Calar Alto high-Resolution search for M dwarfs with Exoearths with Near-infrared and optical Échelle Spectrographs \citep[CARMENES,][]{quirrenbachetal2014} at the 3.5m telescope at the Calar Alto Observatory; and the High Accuracy Radial velocity Planet Searcher \citep[HARPS,][]{mayoretal2003} at the ESO La Silla 3.6m telescope. Details on the data used for each system are discussed in the respective subsections. The complete time series of all spectroscopic data used in this work will be available at the CDS.%are only available at the CDS via anonymous ftp to cdsarc.u-strasbg.fr (TBC) or via TBC.

We derived HARPS-N and HARPS RVs adopting the template-matching approach of the Template-Enhanced Radial velocity Re-analysis Application \citep[TERRA,][]{anglada-escudebutler2012}, which has proven to be more effective than the standard CCF techniques in the analysis of M dwarf spectra \citep{afferetal2016,pergeretal2017}. Moreover, in order to avoid the low-signal-to-noise-ratio (S/N) blue part of the M dwarfs spectra, we computed the TERRA RVs considering only spectral orders redder than $\lambda > 440$ nm \citep{anglada-escudebutler2012}, corresponding to the 22nd and 18th order for HARPS and HARPS-N, respectively.

We derived the stellar parameters of the three targets following the procedure by \citet{maldonadoetal2017}. We also derived the $\log R'_\text{HK}$ following the procedures to extend its definition to M dwarfs \citep[and references therein]{suarezmascarenoetal2016}. All the derived stellar parameters, along with the relevant astrometric and photometric information of the targets are listed in Table \ref{tab:star_par}.

\begin{table*}
\caption{Stellar parameters of the observed targets.}
\small
\label{tab:star_par}
\centering
\begin{tabular}{lccc}
\hline\hline
Parameter & GJ 328 & GJ 649 & GJ 849\\
\hline
\noalign{\smallskip}
Spectral type\tablefootmark{a} & M0.0 & M1.0 & M3.5 \tablefootmark{a} \\
\noalign{\smallskip}
$T_{\text{eff}}$ $[$K$]$\tablefootmark{a} & $3897 \pm 71$ & $3734 \pm 68$ & $3467 \pm 68$  \\
\noalign{\smallskip}
$[$Fe$/$H$]$ $[$dex$]$\tablefootmark{a} & $-0.06 \pm 0.09$ & $-0.15 \pm 0.09$ & $0.09 \pm 0.09$  \\
\noalign{\smallskip}
Mass $[$M$_\odot]$\tablefootmark{a} & $0.65 \pm 0.08$ & $0.51 \pm 0.05$ & $0.45 \pm 0.07$  \\
\noalign{\smallskip}
Radius $[$R$_\odot]$\tablefootmark{a} & $0.63 \pm 0.07$ & $0.50 \pm 0.05$ & $0.45 \pm 0.07$  \\
\noalign{\smallskip}
$\log g$ $[$cgs$]$\tablefootmark{a} & $4.64 \pm 0.07$ & $4.76 \pm 0.04$ & $4.80 \pm 0.07$  \\
\noalign{\smallskip}
$\log \text{L}_*/\text{L}_\odot$\tablefootmark{a} & $-1.08 \pm 0.10$ & $-1.361 \pm 0.087$ & $-1.59 \pm 0.13$  \\
\noalign{\smallskip}
$\log R'_\text{HK}$\tablefootmark{b} & $- 4.679 \pm 0.031$ & $- 4.925 \pm 0.027$ & $- 5.116 \pm 0.031$  \\
\noalign{\smallskip}
\hline                                   %inserts single line
\noalign{\smallskip}
$\alpha$ (J2000)\tablefootmark{c} & 08$^h$:55$^m$:07.6$^s$ & 16$^h$:58$^m$:08.8$^s$ & 22$^h$:09$^m$:40.3$^s$  \\
\noalign{\smallskip}
$\delta$ (J2000)\tablefootmark{c} & +01$^\circ$:32$'$:47.4$''$ & +25$^\circ$:44$'$:39.0$''$ & -04$^\circ$:38$'$:26.7$''$  \\
\noalign{\smallskip}
$B-V$ $[\text{mag}]$ & 1.30 & 1.48 & 1.50 \\
\noalign{\smallskip}
$V$ $[\text{mag}]$ & 9.997 & 9.655 & 10.366  \\
\noalign{\smallskip}
$J$ $[\text{mag}]$ \tablefootmark{d} & 7.191 & 6.448 & 6.510 \\
\noalign{\smallskip}
$H$ $[\text{mag}]$ \tablefootmark{d} & 6.523 & 5.865 & 5.90 \\
\noalign{\smallskip}
$K$ $[\text{mag}]$ \tablefootmark{d} & 6.352 & 5.624 & 5.594 \\
\noalign{\smallskip}
$\pi$ $[\text{mas}]$ \tablefootmark{c} & $48.740 \pm 0.018$ & $	96.233 \pm 0.024$ & $	113.445 \pm 0.030$ \\
\noalign{\smallskip}
$\mu_\alpha$ $[\text{mas yr}^{-1}]$ \tablefootmark{c} & $44.944\pm 0.020$ & $-115.314 \pm 0.021$ & $1132.583 \pm 0.038$ \\
\noalign{\smallskip}
$\mu_\delta$ $[\text{mas yr}^{-1}]$ \tablefootmark{c} & $-1045.876 \pm 0.013$ & $-508.087 \pm 0.026$ & $-22.157 \pm 0.037$ \\
\noalign{\smallskip}
\hline
\end{tabular}
\tablefoot{\tablefoottext{a}{\citet{maldonadoetal2017}}; \tablefoottext{b}{\citet{suarezmascarenoetal2016}};  \tablefoottext{c}{\citet{gaiaetal2022}}; \tablefoottext{d}{\citet{cutrietal2003}}}
\end{table*}

\subsection{GJ 328}
\label{sec:data_gj328}

As part of the GAPS survey,  GJ 328 has been observed from BJD $= 2458213.4$ (4 April 2018) to BJD $= 2459345.4$ (10 May 2021) with the HARPS-N spectrograph. 
The total number of data points acquired was $131$ over a time span of $1132$ days.
The mean internal error of the TERRA  HARPS-N data is $0.97$ m s $^{-1}$, and the r.m.s. is $11.57$ m s $^{-1}$.
We additionally retrieved public RV measurements of GJ 328 from HET/HRS, Keck/HIRES, HJST/Tull, the same used for the detection of GJ 328\,b \citep{robertsonetal2013b}. In addition to these data, we collected public RV measurements collected from OHP/SOPHIE \citep{moultakaetal2004}. The details of the adopted literature time series are listed in Table \ref{tab:data_328}. Combining all the data together, we obtain a time series of 206 RVs, spanning 6700 days, from BJD $= 2452645.8$ (6 January 2003) to BJD $= 2459345.4$ (10 May 2021), with a mean internal error of $\sigma = 3.0$ m s $^{-1}$, and an r.m.s. of $17.4$ m s $^{-1}$, as reported in the final line of Table \ref{tab:data_328}.

\begin{table}
\caption[]{RV time series of GJ328.}
\label{tab:data_328}
\centering
\begin{tabular}{lcccc}
\hline
\hline
Instrument & $N_\text{obs}$ & $T_\text{s}$ & r.m.s. & $\sigma$\\
 & & $[$d$]$ & m s $^{-1}$ & m s $^{-1}$ \\
\hline
\noalign{\smallskip}
HARPS-N & 130 & 1132 & 11.58 & 0.95 \\ 
\noalign{\smallskip}
HRS & 58 & 3752 & 24.5 & 6.4 \\ 
\noalign{\smallskip}
HIRES & 4 & 1092 & 16.5 & 3.6  \\
\noalign{\smallskip}
Tull & 14 & 346 & 9.3 & 5.6 \\
\noalign{\smallskip}
SOPHIE & 36 & 1826 & 22.8 & 4.0 \\
\noalign{\smallskip}
\hline
\noalign{\smallskip}
Combined & 242 & 6700 & 17.4 & 3.0 \\ 
\noalign{\smallskip}
\hline
\end{tabular}
\end{table}

Moreover, in order to measure the stellar activity of GJ328, we computed the stellar activity indexes based on the Ca~{\sc ii}  H and K, H$\alpha$, Na~{\sc i} D$_{\rm 1}$ D$_{\rm 2}$, and He~{\sc i} D$_{\rm 3}$ spectral lines, applying the procedure described in \citet{gomesdasilva11} to all the available HARPS-N spectra.
\citet{robertsonetal2013b} found clues of a long-period activity cycle in the time series of the Na~{\sc i} D$_{\rm 1}$ D$_{\rm 2}$ activity index, and used it to correct the stellar noise in the RV data. For this reason, we collected the activity indexes data from the HRS spectra used in their analysis, paying particular attention to the time series of Na~{\sc i} D$_{\rm 1}$ D$_{\rm 2}$ derived from our HARPS-N spectra. We derived the Na~{\sc i} D$_{\rm 1}$ D$_{\rm 2}$ activity index using both a 0.5 \r{A} window and a 1 \r{A} window, as \citet{robertsonetal2013b} stated that the stellar magnetic behavior of GJ328 was best measured by adopting a 1 \r{A} window, as opposed to the standard recipe \citep{gomesdasilva11}.
We avoided using CCF-based activity indicators, such as BIS or FWHM, since the CCF of M dwarfs is usually heavily distorted and difficult to raliably measure \citep{raineretal2020}.

\subsection{GJ 649}

GJ 649 was observed with HARPS-N from BJD $= 2458213.6$ (5 April 2018) to BJD $= 2459130.3$ (8 October 2020), for a total of 147 spectra over a time span of 917 days. The mean internal error of the TERRA HARPS-N data is $0.76$ m s$^{-1}$, and the r.m.s. is $7.88$ m s $^{-1}$. We collected public Keck/HIRES RV measurements of GJ 649 from the California Legacy Survey catalog \citep{rosenthaletal2021}, as well as public CARMENES RV measurements from CARMENES GTO DR1 \citep{ribasetal2023}. The details of all the adopted RV time series are listed in Table \ref{tab:gj649_data}. Combining the data together, we obtain a time series of 269 RVs, spanning 7720 days, from BJD $= 2451409.8$ (19 August 1999) to BJD $= 2459130.3$ (8 October 2020), with  mean internal error of $\sigma = 1.11$ m s $^{-1}$, and an r.m.s. of $7.75$ m s $^{-1}$, as reported in the final line of Table \ref{tab:gj649_data}.

\begin{table}
\caption[]{RV time series of GJ 649.}
\label{tab:gj649_data}
\centering
\begin{tabular}{lcccc}
\hline
\hline
Instrument & $N_\text{obs}$ & $T_\text{s}$ & r.m.s. & $\sigma$\\
 & & $[$d$]$ & m s $^{-1}$ & m s $^{-1}$ \\
\hline
\noalign{\smallskip}
HARPS-N & 147 & 917 & 7.88 & 0.76 \\ 
\noalign{\smallskip}
HIRES-pre & 24 & 1786 & 8.04 & 1.64  \\
\noalign{\smallskip}
HIRES-post & 44 & 4960 & 9.12 & 1.21  \\
\noalign{\smallskip}
CARMENES & 54 & 399 & 5.00 & 1.77  \\
\noalign{\smallskip}
\hline
\noalign{\smallskip}
Combined & 269 & 7720 & 7.75 & 1.11 \\ 
\noalign{\smallskip}
\hline
\end{tabular}
\end{table}

We also computed the stellar Ca~{\sc ii}  H and K, H$\alpha$, Na~{\sc i} D$_{\rm 1}$ D$_{\rm 2}$, and He~{\sc i} D$_{\rm 3}$ activity indexes from the HARPS-N spectra, as detailed in the previous section.

\subsection{GJ 849}

GJ 849 was observed with HARPS-N from BJD $= 2458269.7$ (31 May 2018) to BJD $= 2459213.3$ (29 December 2020), for a total of 94 spectra over a time span of 944 days. The mean internal error of the TERRA HARPS-N data is $0.81$ m s$^{-1}$, and the r.m.s. is $13.50$ m s $^{-1}$. We collected public Keck/HIRES RV measurements of GJ 849 from the California Legacy Survey catalog \citep{rosenthaletal2021}, and public HARPS spectra from the ESO archive (ESO programmes 072.C-0488(E) and 183.C-0437(A)), from which we computed the RVs using TERRA. Moreover, we collected public CARMENES RV measurements from CARMENES GTO DR1 \citep{ribasetal2023}. The details of all the adopted RV time series are listed in Table \ref{tab:gj849_data}. Combining the data together, we obtain a time series of 323 RVs, spanning 8607 days, from BJD $= 2450606.1$ (6 June 1997) to BJD $= 2459213.3$ (29 December 2020), with  mean internal error of $1.39$ m s $^{-1}$, and an r.m.s. of $17.87$ m s $^{-1}$, as reported in the final line of Table \ref{tab:gj849_data}.

\begin{table}
\caption[]{RV time series of GJ 849.}
\label{tab:gj849_data}
\centering
\begin{tabular}{lcccc}
\hline
\hline
Instrument & $N_\text{obs}$ & $T_\text{s}$ & r.m.s. & $\sigma$\\
 & & $[$d$]$ & m s $^{-1}$ & m s $^{-1}$ \\
\hline
\noalign{\smallskip}
HARPS-N & 94 & 944 & 13.50 & 0.81 \\ 
\noalign{\smallskip}
HIRES-pre & 24 & 2591 & 14.78 & 3.19  \\
\noalign{\smallskip}
HIRES-post & 97 & 5423 & 20.04 & 1.54  \\
\noalign{\smallskip}
HARPS & 48 & 2824 & 19.29 & 1.03 \\ 
\noalign{\smallskip}
CARMENES & 60 & 1250 & 17.00 & 1.65 \\ 
\noalign{\smallskip}
\hline
\noalign{\smallskip}
Combined & 323 & 8607 & 17.87 & 1.39 \\ 
\noalign{\smallskip}
\hline
\end{tabular}
\end{table}

We also computed the stellar Ca~{\sc ii}  H and K, H$\alpha$, Na~{\sc i} D$_{\rm 1}$ D$_{\rm 2}$, and He~{\sc i} D$_{\rm 3}$ activity indexes from the HARPS-N and HARPS spectra, as detailed in Sect. \ref{sec:data_gj328}.

\section{Spectroscopic time-series analyses}
\label{sec:spectr_analyses}

In this section, we discuss in detail the analyses of the RV and activity time series of the three targets of the sample. In all analyses, the identification of periodic signals in a time series is performed via Generalized Lomb-Scargle periodogram \citep[GLS,][]{zechkur2009}. The complete fits of stellar and planetary signals are instead performed via Markov Chain Monte Carlo (MCMC): this technique was applied via the \texttt{emcee} Affine Invariant MCMC Ensemble sampler by \citet{foreman13}.
We modelled the RV time series with multi-Keplerian models, depending on the number of planets present in each system, taking into account linear RV trends when necessary, and separate offset and jitter terms for each of data-collecting instruments. This general model can be expressed by the equation:
\begin{equation}
\label{eq:Kepl_RV_model}
 \Delta \text{RV} (t) = \gamma_\text{instr} + \sum_{j=1}^{N_p} \Delta \text{RV}_{\text{Kep},j} (t),
\end{equation}
where $\gamma_\text{instr}$ represents each instrument RV offset, $\bar{t}$ is the mean epoch of the time series, $N_p$ is the number of planets in the system, and $\Delta \text{RV}_{\text{Kep},j} (t)$ is the Keplerian RV term of the $j$th planet as a function of the orbital parameters:
\begin{equation}
\label{eq:rv_kepl}
\Delta RV_\text{Kep}(t) =  K [\cos(\nu(t,e,T_{0},P) + \omega) + e \cos(\omega)] .
\end{equation}
In addition to the general model in Eq. \ref{eq:Kepl_RV_model}, we tested an additional  long-term RV acceleration component, $d (t-\bar{t})$, to inquire the presence of possible long-term trends due to stellar cycles and very-long-period additional companions. This term would be added to the model whenever statistically favored. To compare different models and select the statistically favored, we adopt the Bayasian Information Criterion \citep[BIC,][]{schwarz1978}.

Moreover, in order to model and subtract stellar activity signals in the RV data, when present, we employ the Gaussian Process (GP) regression technique. We adopt the most common kernel to model stellar activity signals \citep[e.g.][]{haywoodetal2014,afferetal2016,pinamontietal2019}, i.e. the Quasi-Periodic (QP) kernel:

\begin{equation}
\label{eq:gp_kernel}
K_{i,j} = h^2 \cdot \exp \bigg[ -\frac{(t_i-t_j)^2}{2 \lambda^2} - \frac{\sin^{2}(\pi(t_i-t_j)/\theta)}{2 w^2}\bigg] + \sigma^{2} \cdot \delta_{i,j} ,
\end{equation}
where $K_{i,j}$ is the $i j$ element of the covariance matrix, and the covariance is described by four hyper-parameters: $h$ is the amplitude of the correlation, $\lambda$ is the timescale of decay of the exponential correlation, $\theta$ is the period of the periodic component, and $w$ is the weight of the periodic component. The last term of Eq. \ref{eq:gp_kernel} describes the white noise component of the covariance matrix, where $\delta_{i,j}$ is the Kronecker delta and $\sigma$ is the RV total uncertainty defined as: $\sigma^{2} = \sigma_\text{data}^{2} +  \sigma_\text{jit}^{2}$, with  $\sigma_\text{data}$ the internal error of the data, and  $\sigma_\text{jit}$ the uncorrelated jitter fitted by the MCMC model.

Due to the HIRES CCD upgrade that occurred in August 2004, whenever HIRES data were used in the RV analyses, we considered the data before and after the upgrade as two independent datasets. For each HIRES dataset, we computed independent offset and jitter terms, to avoid zero-point errors \citep[and references therein]{pinamontietal2018}.

\subsection{GJ 328}
\label{sec:spectr_gj328}

As a first test in the study of the GJ 328 system, we used our combined RV time series to fit the known planet's signal, in order to see how the newly taken HARPS-N observations affect its orbital parameters.
We modelled the combined datasets with a single-Keplerian model, as specified by Eq. \ref{eq:Kepl_RV_model} with $N_p = 1$.

\begin{figure}
   \centering
   \subfloat[][]
 {\includegraphics[width=.45\textwidth]{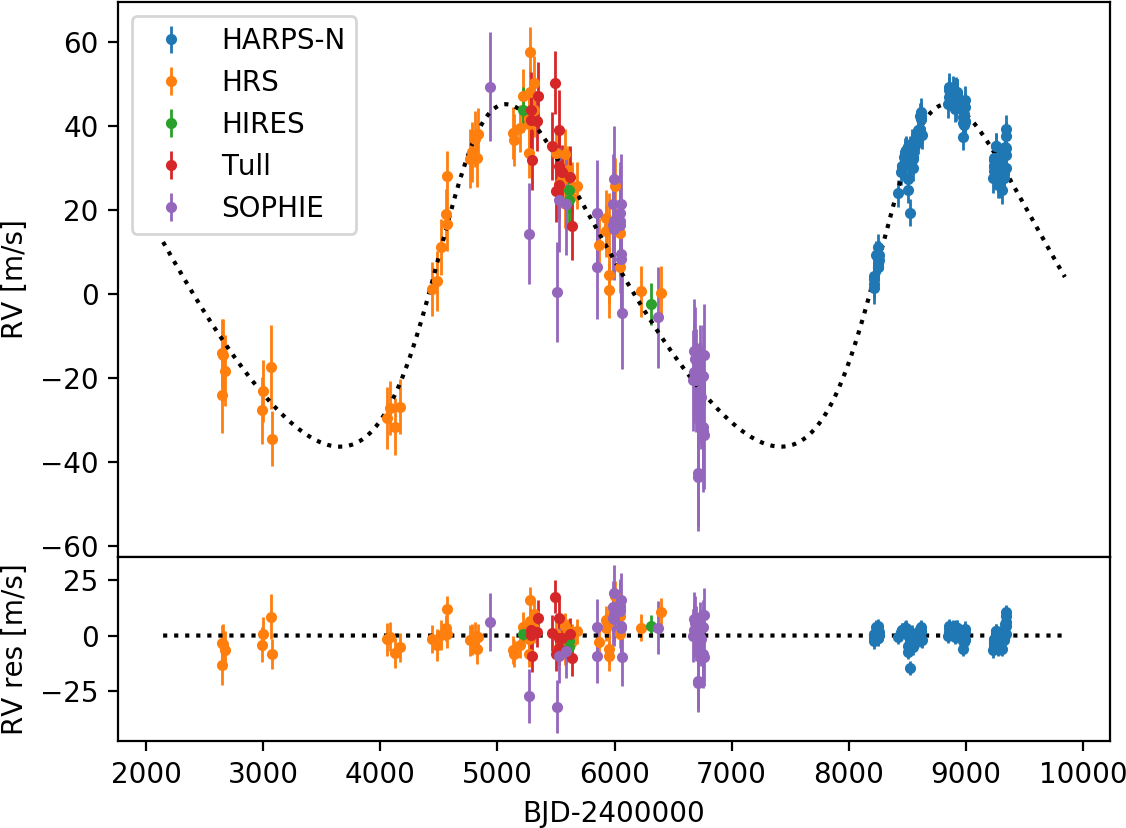}
 \label{fig:gj328_oneplan_mod}} \\
   \subfloat[][]
 {\includegraphics[width=.45\textwidth]{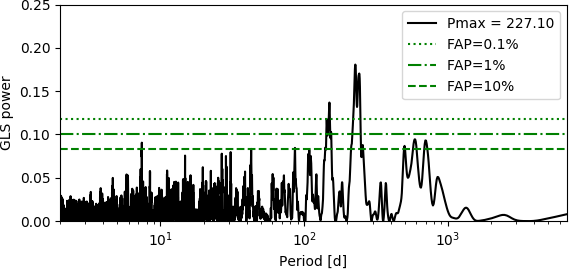}
 \label{fig:gj328_oneplan_residuals}}
 \caption{GJ 328: Best-fit RV one-planet model, corrected for the instrumental offsets and RV residuals (top panel), GLS periodogram of the RV residuals (bottom panel).}
 \label{fig:gj328_oneplan}
\end{figure}

\begin{figure}
   \centering
 \includegraphics[width=.45\textwidth]{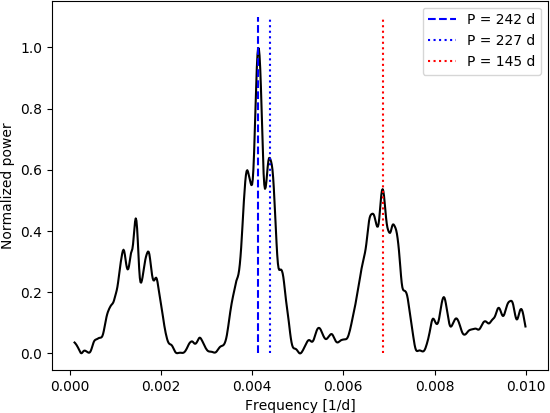}
 \caption{GJ 328: window function GLS periodogram structure of the $P=242$ d signal.}
 \label{fig:gj328_aliases}
\end{figure}

\begin{figure*}
   \centering
   \subfloat[][\emph{Ca~{\sc ii}  H \& K}]
   {\includegraphics[width=.45\textwidth]{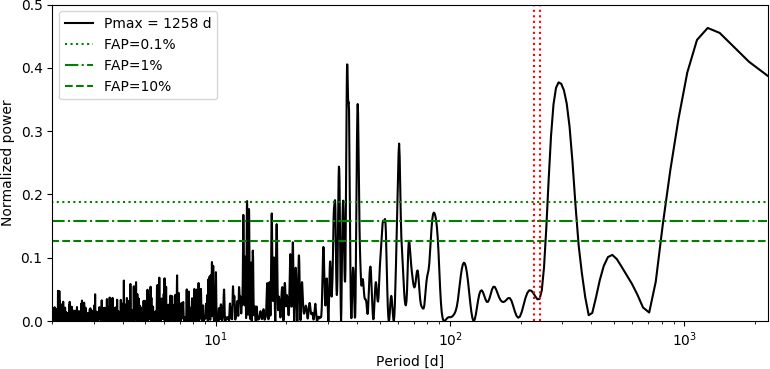}} \,
   \subfloat[][\emph{H$\alpha$}]
   {\includegraphics[width=.45\textwidth]{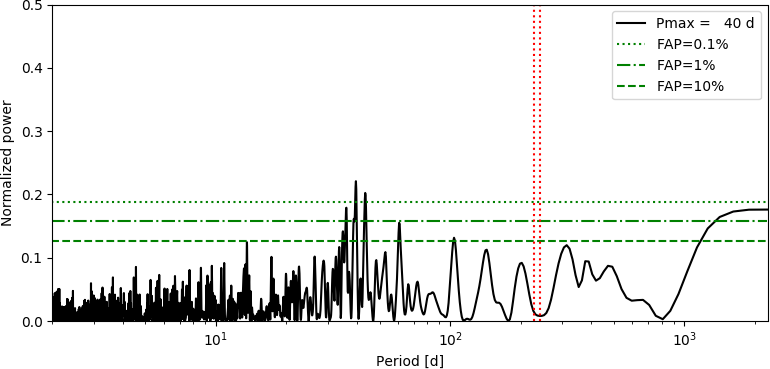}} \\
   \subfloat[][\emph{Na~{\sc i} D$_{\rm 1}$ D$_{\rm 2}$}]
   {\includegraphics[width=.45\textwidth]{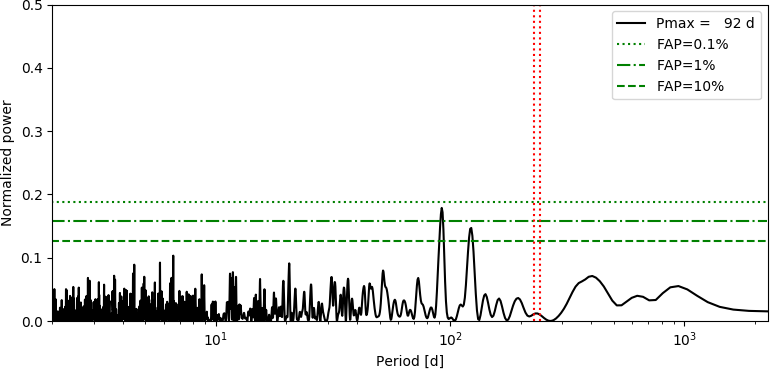}} \,
   \subfloat[][\emph{He~{\sc i} D$_{\rm 3}$}]
   {\includegraphics[width=.45\textwidth]{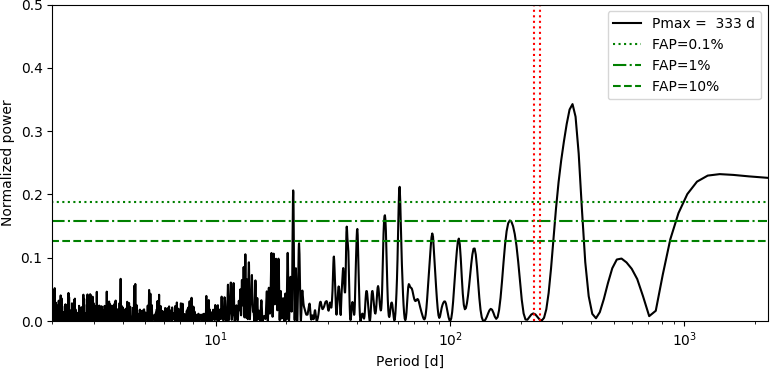}}
      \caption{GJ 328: GLS periodograms of the HARPS-N activity indexes computed following the procedure by \citet{gomesdasilva11}.
      The red dotted vertical lines indicate the 227-242 d period of the signals identified in the RV residuals. The horizontal lines indicate the FAP levels as in Fig. \ref{fig:gj328_oneplan_residuals}.}
         \label{fig:gj328_activity_periodograms}
\end{figure*}

From this 1-Keplerian model, we obtained an amplitude of  $K_b = 40.8^{+1.7}_{-1.6}$ m s$^{-1}$ and a period of $P_b = 3763^{+17}_{-17}$ d, in agreement ($\sim 1\sigma$) with the values reported by \citet{robertsonetal2013b}. The complete resulting best-fit parameters of this model are listed in Table \ref{tab:gj328_mcmc_prior_param}, and the best-fit model is shown Fig. \ref{fig:gj328_oneplan_mod}, after correcting for the instrumental offsets. We tested the addition of a linear trend to the model, but the resulting acceleration was not strongly significant, $d = 0.0026^{+0.0012}_{-0.0011}$ m s$^{-1}$ d$^{-1}$ i.e. $2.4\sigma$, it is strongly correlated with the HARPS-N offset, $\gamma_\text{HARPS-N}$, and does not increase the BIC, $\Delta\text{BIC} \simeq 0$. In Fig. \ref{fig:gj328_oneplan_residuals} we can see the GLS periodogram of the RV residuals after subtracting the one-planet model, which is dominated by a significant peak, False Alarm Probability (FAP) $<< 0.1\%$, at around 227 d, with a strong alias at 242 d. Fig. \ref{fig:gj328_aliases} shows the window function structure of the $P = 242$ d peak, which shows a strong alias at $\Delta f = 0.00025$ d$^{-1}$ corresponding to $P = 227$ d, caused by the large gap between the SOPHIE and HARPS-N observations. We can also see the 1-year alias at $\Delta f = 0.0027$ d$^{-1}$ corresponding to $P = 145$ d. This signal was not previously reported in the original analysis by \citet{robertsonetal2013b}, and thus we check the HARPS-N activity indicators time series of GJ 328 to ascertain its nature: Fig. \ref{fig:gj328_activity_periodograms} shows the GLS periodogram of the activity indexes. The \emph{Ca~{\sc ii}} and \emph{H$\alpha$} periodograms show prominent peaks at around 35-40 d, which are probably related to the rotation period of the star.
Applying the activity–rotation relationships for early-M dwarfs by \citet{suarezmascarenoetal2018}, we obtain an estimate of $P_\text{rot} = 19^{+24}_{-11}$ d, which is compatible with the observed peaks in the \emph{Ca~{\sc ii}} and \emph{H$\alpha$} periodograms. Moreover, the \emph{Ca~{\sc ii}} time-series is dominated by strong long-term periodicities at $\simeq 300$ d and $\simeq 1300$ d; this additional periodic signals may be related to a long-term magnetic cycle of the star, however it is difficult to confirm as the current HARPS-N data cover less than one full 1300 d cycle.
We tested for correlation between the HARPS-N activity time-series and the RVs, and found no significant correlation ($\left | \rho \right | < 0.3$).
Finally, we can see, no significant peak is found near the 227-242 d periods identified in the RVs.

\begin{figure}
   \centering
 \includegraphics[width=.45\textwidth]{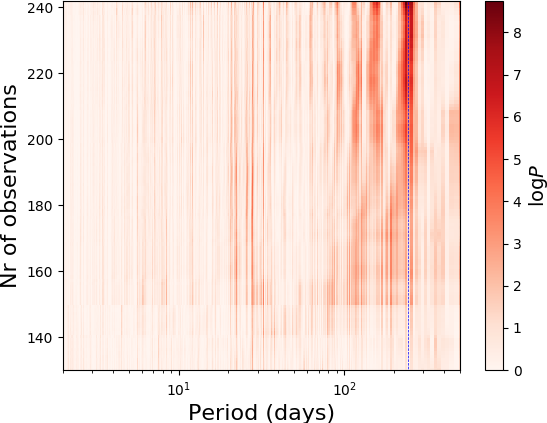}
 \caption{GJ 328: Stacked BGLS of the RV residuals after subtracting GJ 328\,b model. The blue dashed line mark the 242 d periodicity identified in the data.}
 \label{fig:gj328_sbgls}
\end{figure}

We then tested the coherency of the 227-242 d periodicity over time, by means of a Stacked Bayesian GLS \citep[BGLS,][]{mortieretal2017}, as Keplerian signals should always increase in strenght, while stellar signals grow stronger and weaker over time due to their evolving nature. Fig. \ref{fig:gj328_sbgls} shows the Stacked BGLS of GJ 328 RVs after subtracting the long-period planet model, starting from the beginning of HARPS-N observations: we can see how the strenght of the peaks at 227-242 d increases steadily throughout the observations.

\begin{figure}
   \centering
   \subfloat[][]
 {\includegraphics[width=.45\textwidth]{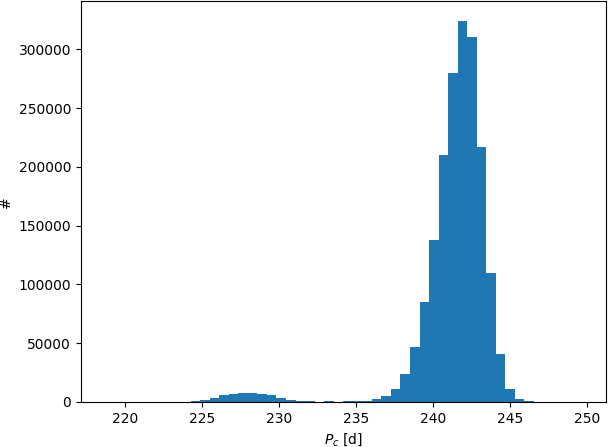}
 \label{fig:gj328_242vs227:hist}} \\
   \subfloat[][]
 {\includegraphics[width=.45\textwidth]{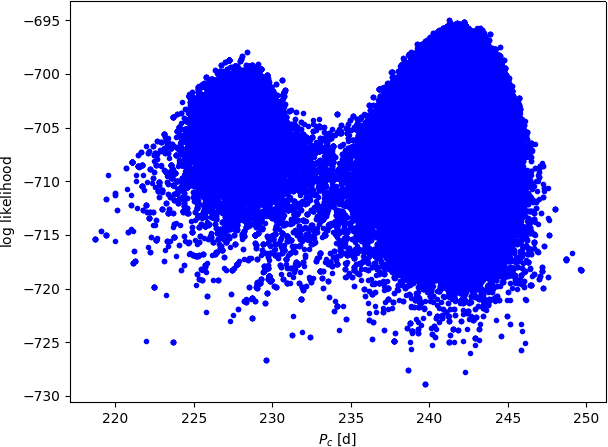}
 \label{fig:gj328_242vs227:like}}
 \caption{GJ 328: comparison of the $P_c = 227$ d and $P_c = 242$ d signals in the two-Keplerian model, shown as the histogram of the posterior distribution of $P_c$ (top panel), and the log-likelihood as a function of $P_c$ (bottom panel).}
 \label{fig:gj328_242vs227}
\end{figure}

We then proceeded to model the RV time series with a two-Keplerian model, taking into account both the known 3700 d planet and the new 227-242 d signal. This was done by adding another Keplerian signal $\Delta RV_\text{Kep,2}(t)$ to Eq. \ref{eq:rv_kepl}. The model favored the 242 d period as the best-fit of the data, with $P_c = 241.8^{+1.3}_{-1.7}$ d and $K_c = 2.95^{+0.39}_{-0.38}$ m s$^{-1}$. Fig. \ref{fig:gj328_242vs227} shows how the MCMC fit strongly favoured the $P_c = 242$ d solution in the posterior distribution, due to its higher likelihood. We thus adopted that as the true signal's period, considering 227 d to be its alias. To test the significance of the eccentricity of this new signal, we performed two separate analyses, one fixing the eccentricity to $e_c = 0$ and one leaving it free. All the best-fit parameters of these models are listed in Tab. \ref{tab:gj328_mcmc_prior_param}.
The addition of a second signal to the model produces a significant decrease of the BIC, $\Delta \text{BIC} \simeq -40$, confirming the presence of the additional signal in the RV data. Moreover, there is a significant improvement in the BIC of the tested two-Keplerian  $e_c = 0$ model ($\Delta \text{BIC} \simeq -8$) and we thus adopted it as the best model to describe GJ 328 time series. The computed eccentricity in the eccentric model, $e_c = 0.22^{+0.15}_{-0.13}$, is not significant ($1.7\sigma$), and corresponds to an upper limit of $e_c < 0.28$ from the 68th percentile of the posterior distribution.

\begin{figure*}
   \centering
   \subfloat[][]
   {\includegraphics[width=.45\textwidth]{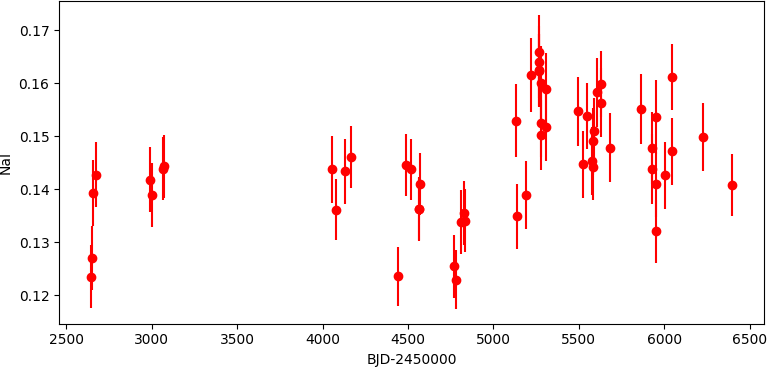}} \,
   \subfloat[][]
   {\includegraphics[width=.45\textwidth]{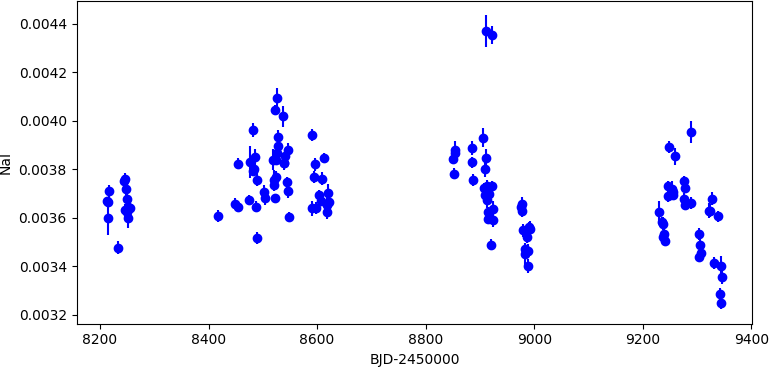}} \\
   \subfloat[][]
   {\includegraphics[width=.45\textwidth]{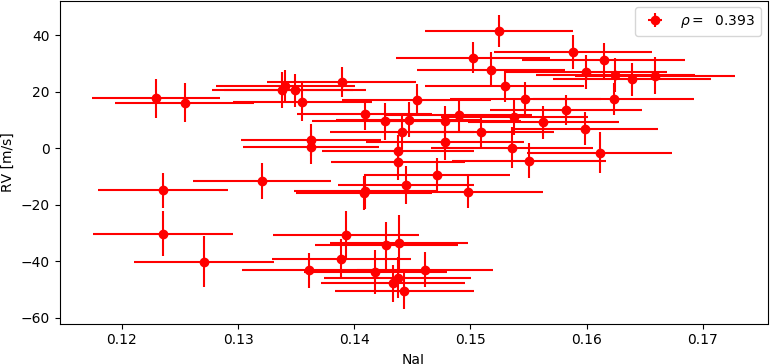}} \,
   \subfloat[][]
   {\includegraphics[width=.43\textwidth]{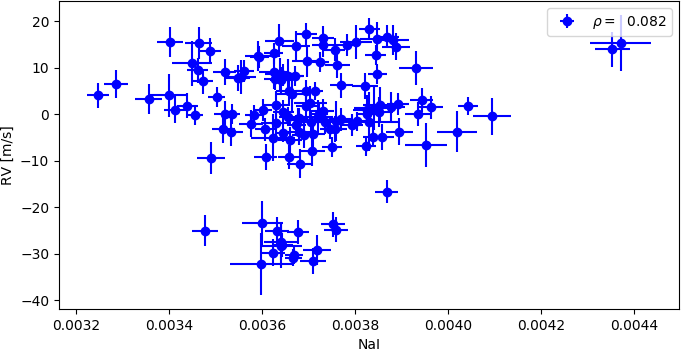}}
      \caption{GJ 328: time series of the Na~{\sc i} D$_{\rm 1}$ D$_{\rm 2}$ activity index computed from HRS (upper left) and HARPS-N (upper right). Correlation between the RV and Na~{\sc i} activity index computed for HRS (lower left) and HARPS-N (lower right) data.}
         \label{fig:gj328_NaI_activity}
\end{figure*}

\citet{robertsonetal2013b} suggested the presence of a long-period magnetic cycle in GJ 328 activity data: the authors found a 2000 d periodicity in the Na~{\sc i} D$_{\rm 1}$ D$_{\rm 2}$ HRS time series, measuring a $\rho = 0.41$ correlation with the HRS RVs, and then proceeded to correct the influence of the magnetic cycles via a linear fit of the RV-activity dependence. We tested the presence of this magnetic cycle, both in our newly acquired HARPS-N data and in the archival spectroscopic data. The Na~{\sc i} activity index was not available for HIRES and Tull data, so we studied only the HRS and HARPS-N data. Figure \ref{fig:gj328_NaI_activity} shows the Na~{\sc i} time series and correlation with the respective RVs for the two instruments. As shown in the bottom part of the Figure, we measure a Pearson correlation coefficient of $\rho = 0.39$ and $\rho = 0.08$ for HRS and HARPS-N data, respectively. We find no evidence of a strong correlation, in particular in the HARPS-N data. We tested also the Na~{\sc i} activity index computed from HARPS-N spectra following \citet{robertsonetal2013b} recipe, finding again no significant correlation, $\rho = -0.03$. Moreover, as shown in Fig. \ref{fig:gj328_activity_periodograms}, the HARPS-N Na~{\sc i} time series does not show any long-term periodic signal. We thus find no evidence of the 2000 d magnetic cycle reported by \citet{robertsonetal2013b}, and decided not to take it into account in our modeling of GJ 328's planetary signals. As an additional precaution, we reproduced our one-Keplerian model of the RV data after correcting the HRS RVs following the linear fit with Na~{\sc i} proposed in \citet{robertsonetal2013b}, and found no significant difference ($< 1\sigma$) in the resulting planetary parameters.

As an additional test on the presence of stellar noise in the RV time-series, we performed an additional mcmc fit including a GP component, to model possibile stellar signals in the data. The resulting amplitude of the stellar GP component was consistent with 0, and the model was statistically disfavored ($\Delta\text{BIC} \simeq +10$). We thus see no evidence of stellar signals in the RV data.

\begin{figure}
   \centering
   \subfloat[][]
 {\includegraphics[width=.45\textwidth]{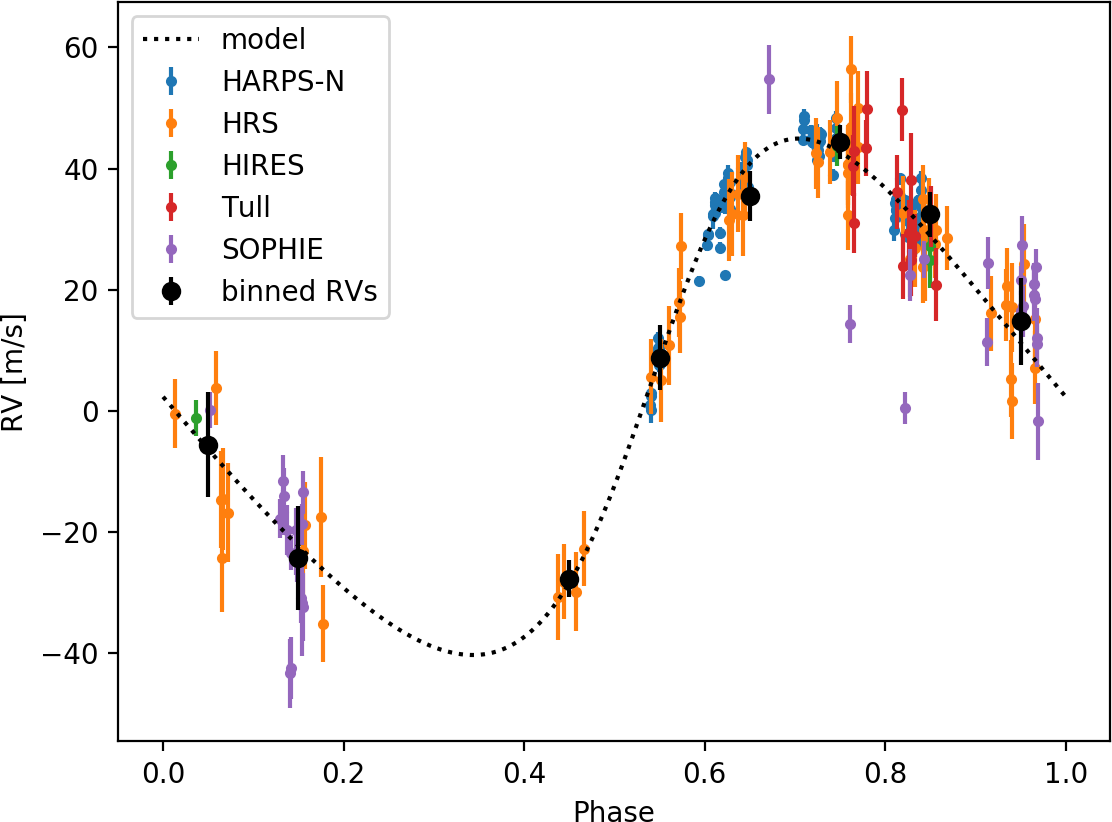}
 \label{fig:gj328_phasefolded1}} \\
   \subfloat[][]
 {\includegraphics[width=.45\textwidth]{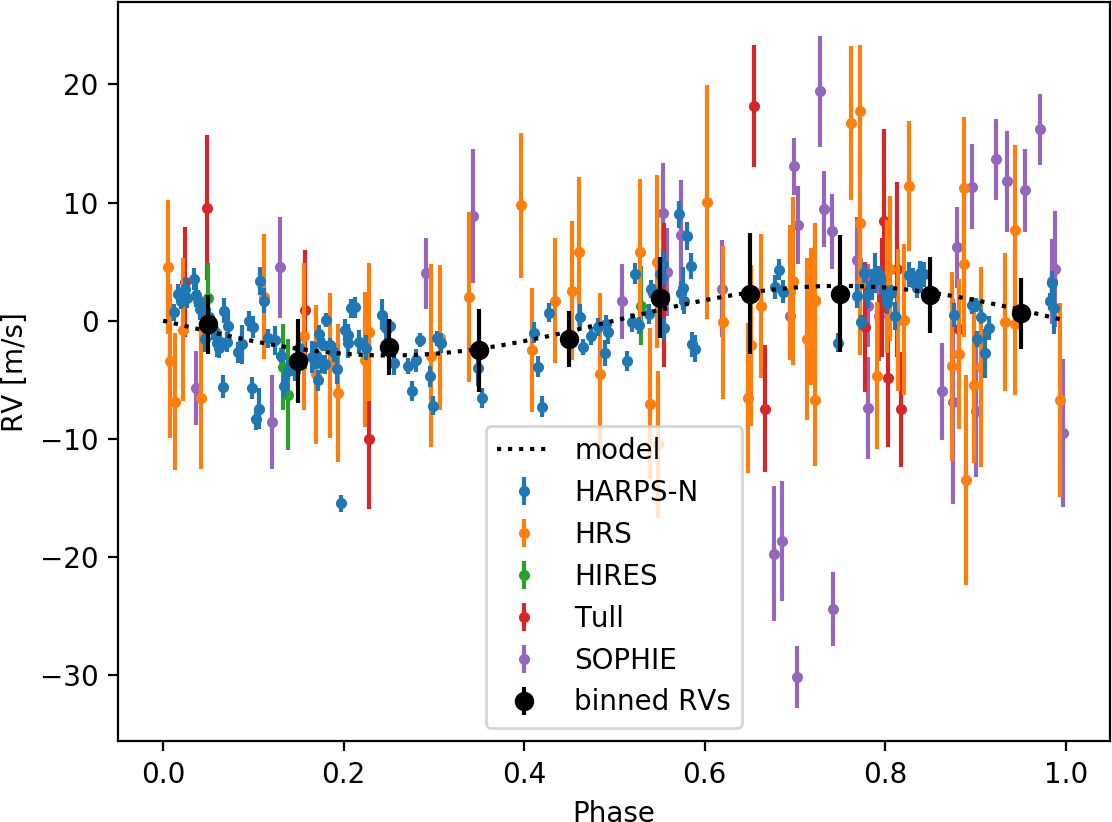}
 \label{fig:gj328_phasefolded2}}
 \caption{GJ 328: Phase-folded RV signal of GJ 328\,b (\textit{Upper panel}) and GJ 328\,c (\textit{Lower panel}). The black points and error bars represent the binned weighted averages and standard deviations of the data.}
 \label{fig:gj328_phasefolded}
\end{figure}

We thus confirm that the 242 d RV signal is best explained by the presence of an additional planetary companion, hereafter GJ 328\,c, a sub-Neptune of minimum-mass $m_c \sin i = 21.4^{+ 3.4}_{- 3.2}$ M$_\oplus$. The best-fit orbital parameters of the planetary signals in the final mcmc model are listed in Tab. \ref{tab:all_orbital_parameters}. The phase-folded RV curves of the two planetary signals are shown in Fig. \ref{fig:gj328_phasefolded}. It is worth noticing in Fig. \ref{fig:gj328_phasefolded2} that the model of GJ 328\,c follows closely the HARPS-N RV data, while the other time series are widely spread due to the large uncertainties and jitters (see Tab. \ref{tab:gj328_mcmc_prior_param}).
No significant signal was identified in the GLS periodogram of the RV residuals of the 2-Keplerian model.

\subsection{GJ 649}
\label{sec:spectr_gj649}

\begin{figure}
   \centering
   \subfloat[][]
 {\includegraphics[width=.45\textwidth]{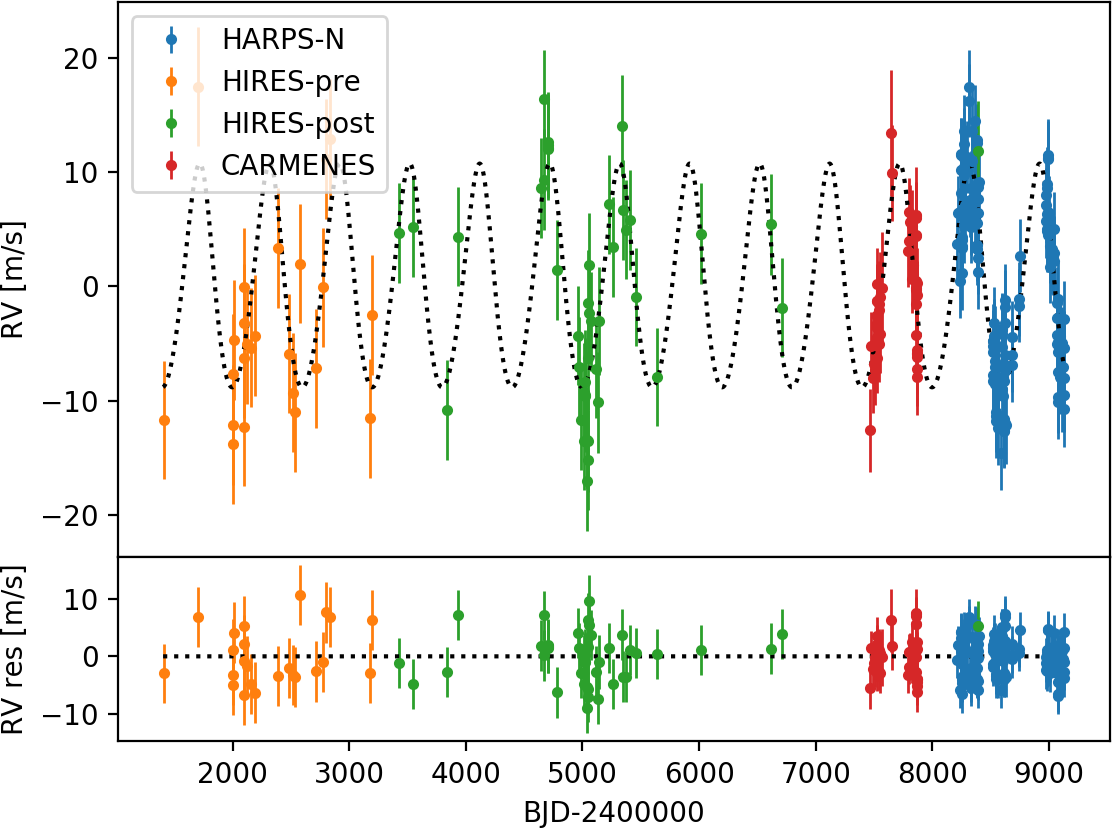}
 \label{fig:gj649_oneplan_mod}} \\
   \subfloat[][]
 {\includegraphics[width=.45\textwidth]{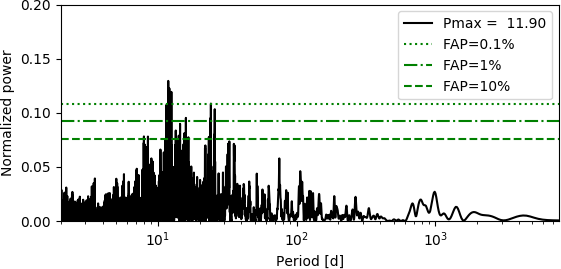}
 \label{fig:gj649_oneplan_residuals}}
 \caption{GJ 649: Best-fit RV one-planet model, corrected for the instrumental offsets, and RV residuals (top panel), and GLS periodogram of the RV residuals (bottom panel).}
 \label{fig:gj649_oneplan}
\end{figure}

At first, we used our combined RV time series to test and update the one-Keplerian model corresponding to the known planet GJ 649\,b signal, following Eq. \ref{eq:Kepl_RV_model} and \ref{eq:rv_kepl} as previously discussed. The best-fit model and GLS periodogram of the RV residuals are shown in Fig. \ref{fig:gj649_oneplan_mod} and \ref{fig:gj649_oneplan_residuals} respectively, while the complete set of adopted priors and best-fit parameters is listed in Table \ref{tab:gj649_mcmc_prior_param}. We did not include an acceleration term, $d$, in Eq. \ref{eq:Kepl_RV_model} as its inclusion in the model decreased the statistical significance of the model (higher BIC) and resulted in a best-fit acceleration compatible with zero ($<2\sigma$). From this fit we obtain an amplitude of $K_b = 9.8^{+0.3}_{-0.3}$ m s$^{-1}$, a period of $P_b = 600.1^{+1.5}_{-1.6}$ d, and an eccentricity of $0.114^{+0.041}_{-0.041}$:  it is worth noticing that, while the measured period is well compatible within $1\sigma$ with the value reported by \citet{johnsonetal2010}, the amplitude and the eccentricity are quite smaller ($\simeq 2\sigma$) than their values. Moreover, as we can see in Fig. \ref{fig:gj649_oneplan_residuals}, the periodogram of the residuals shows a significant peak at around 12 d. This period is close to half the stellar rotation period measured by \citet{johnsonetal2010}, $P_\text{rot} = 24.8 \pm 1.0$ d, and later confirmed by \citet{diezalonsoetal2019}, $P_\text{rot} = 23.8 \pm 0.1$ d. This leads us to suspect it to be in fact an harmonic of the stellar rotation period, caused by chromospheric activity. This stellar rotation period is confirmed by the GLS analyses of the HARPS-N activity indicators (not shown), which identify strong $\sim 24$ d periodicities in the Ca~{\sc ii} and H$\alpha$ time series.

\begin{figure}
   \centering
   \subfloat[][]
 {\includegraphics[width=.45\textwidth]{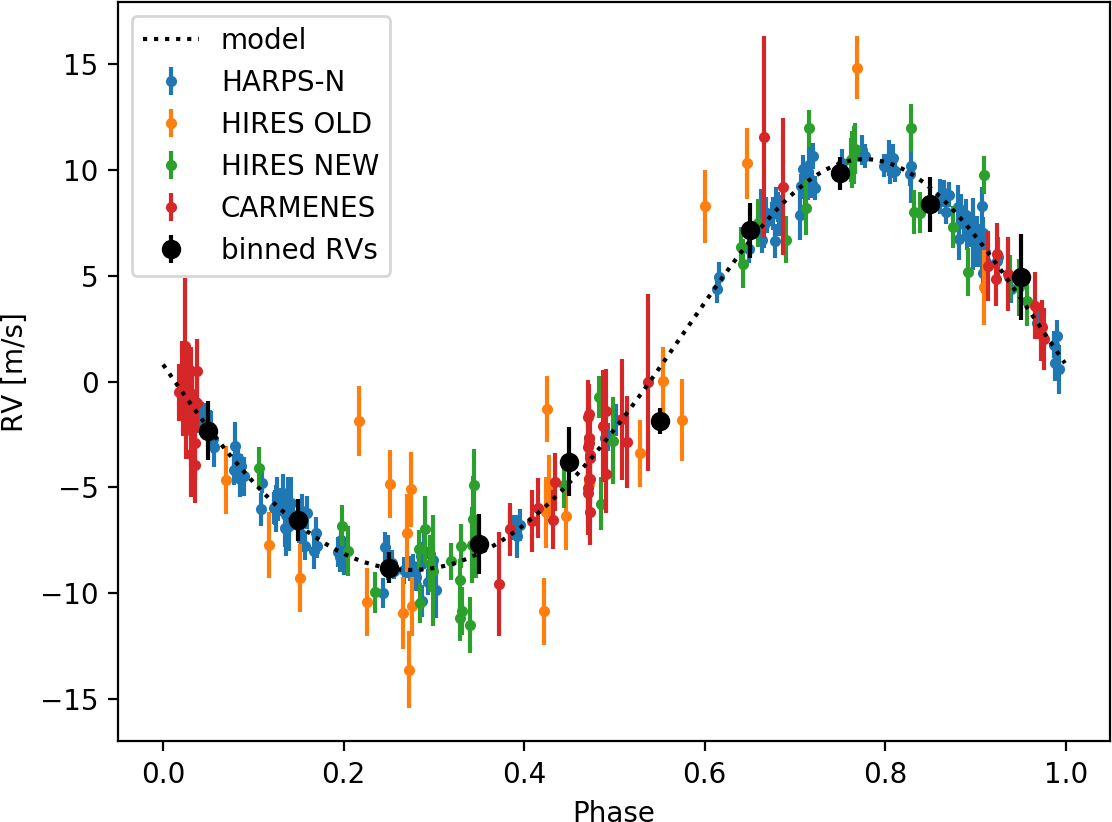}
 \label{fig:gj649_phasefolded}} \\
   \subfloat[][]
 {\includegraphics[width=.45\textwidth]{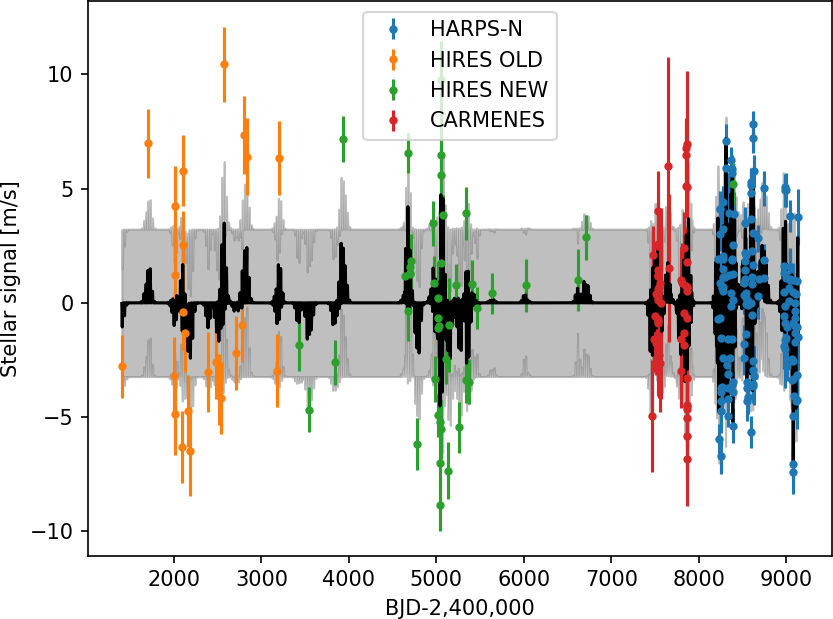}
 \label{fig:gj649_gpmodel}}
 \caption{GJ 649: \textit{Upper panel:} Phase-folded RV signal of GJ 649\,b (as in Fig. \ref{fig:gj328_phasefolded}), after the subtraction of the stellar correlated signal. \textit{Lower panel:} best-fit stellar quasi-periodic signal obtained from the GP + 1 planet model (black line) compared to the RV residuals.}
 \label{fig:gj649_final}
\end{figure}

To confirm the stellar origin of the 12 d RV signal, we applied the GP regression as described in Eq. \ref{eq:gp_kernel}. As listed in Tab. \ref{tab:gj649_mcmc_prior_param}, we adopted a broad uniform prior for the hyper-parameter $\theta$, which corresponds to the stellar rotation period, $\mathcal{U}$(10,50) d, in order to include both the identified periodicity of 12 d and the expected stellar rotation period of 24 d. The GP model converged to $\theta = 24.89^{+0.34}_{-0.35}$ d, and the 12 d signal disappeared from the RV residuals, proving it was in fact produced by the chromospheric activity of the target. Our measured value of $\theta$ confirms the 24 d rotation period measured in the literature. It is also worth noticing that, opposite to the claim by \citet{johnsonetal2010}, we do not find a significant orbital eccentricity in our final model ($e = 0.083^{+0.068}_{-0.055}$), which could mean that the apparent eccentricity measured in their fit was due to the sub-optimal sampling of the original RV time series, as well as to the lack of stellar activity correction. The other orbital parameters do not vary significantly between the two fits with and without GP modeling of the stellar activity (see Tab. \ref{tab:gj649_mcmc_prior_param}). All the best-fit orbital parameters of the planetary signals in the final mcmc model are listed in Tab. \ref{tab:all_orbital_parameters}. The complete details of the model are shown in Appendix \ref{app:emcee}. The phase-folded RV signal of GJ 649\,b and quasi-periodic stellar model obtained from the simultaneous GP + 1 planet fit are shown in Fig. \ref{fig:gj649_phasefolded} and \ref{fig:gj649_gpmodel}, respectively. No additional signal is found in GLS periodograms of the residuals, and thus we adopt the Keplerian + GP model as the best fit to GJ 649 RV time series.

\subsection{GJ 849}
\label{sec:spectr_gj849}

To recover and update the orbital RV signals of GJ 849\,b and c, we fitted a 2-Keplerian model via our \texttt{emcee} setup on the combined RV time series of GJ 849, expressed by Eq. \ref{eq:Kepl_RV_model} with $N_p = 2$. As in the previous Section, we did not include an acceleration term, $d$, in the RV model, as it resulted in a higher BIC and a best-fit acceleration $\simeq 0$.

\begin{figure}
   \centering
   \subfloat[][]
 {\includegraphics[width=.45\textwidth]{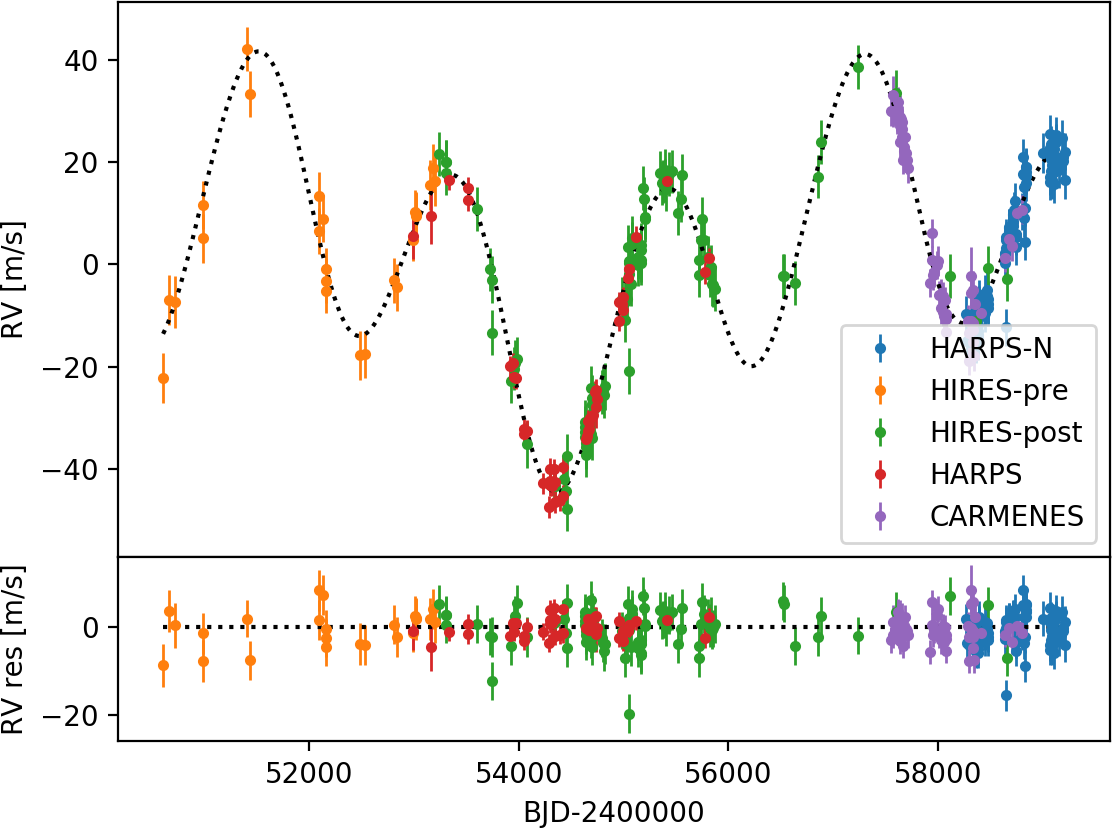}
 \label{fig:gj849_twoplan_mod}} \\
   \subfloat[][]
 {\includegraphics[width=.45\textwidth]{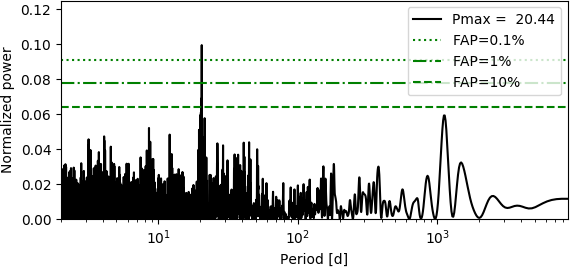}
 \label{fig:gj849_twoplan_residuals}}
 \caption{GJ 849: \textit{Upper panel:} best-fit 2-Keplerian model of GJ 849\,b and c, corrected for the instrumental offsets, and RV residuals; \textit{Lower panel:} GLS periodogram of the RV residuals.}
 \label{fig:gj849_twoplan}
\end{figure}

The best-fit model and GLS periodogram of the RV residuals are shown in Fig. \ref{fig:gj849_twoplan_mod} and \ref{fig:gj849_twoplan_residuals}, respectively, while the complete set of adopted priors and best-fit parameters is listed in Table \ref{tab:gj849_mcmc_prior_param}. Our results confirm within the error bars the minimum masses computed by \citet{fengetal2015} for the two planets, as well as the orbital period of GJ 849\,b, while we find a marginally longer orbital period for GJ849\,c with respect to \citet{fengetal2015} ($<2\sigma$). Moreover, we confirm the low values of the orbital eccentricities of the two planets, and thus also the dynamical stability of the system \citep{fengetal2015}.

The residuals show a significant periodic signal at $P = 20$ d (see Fig. \ref{fig:gj849_twoplan_residuals}). This short-period signal could be related to the stellar rotation period of GJ 849: \citet{suarezmascarenoetal2015} measured a rotation period of $P_\text{rot} = 39.2 \pm 6.3$ d from the activity indicators derived from 42 HARPS spectra, and the periodic signal we observe in the RV residuals is close to $P_\text{rot}/2$.
We tested this by analysing the activity indices derived from all the available HARPS and HARPS-N spectra. The GLS periodogram of both CaII H$\&$K and H$\alpha$ time series (not shown) are dominated by periodic signals at both $\simeq 22$ d and $\simeq 40$ d, which support the measured $P_\text{rot}$ by \citet{suarezmascarenoetal2015}, and confirm the presence of a strong harmonic signal at $P_\text{rot}/2$.

\begin{figure}
   \centering
   \subfloat[][]
 {\includegraphics[width=.45\textwidth]{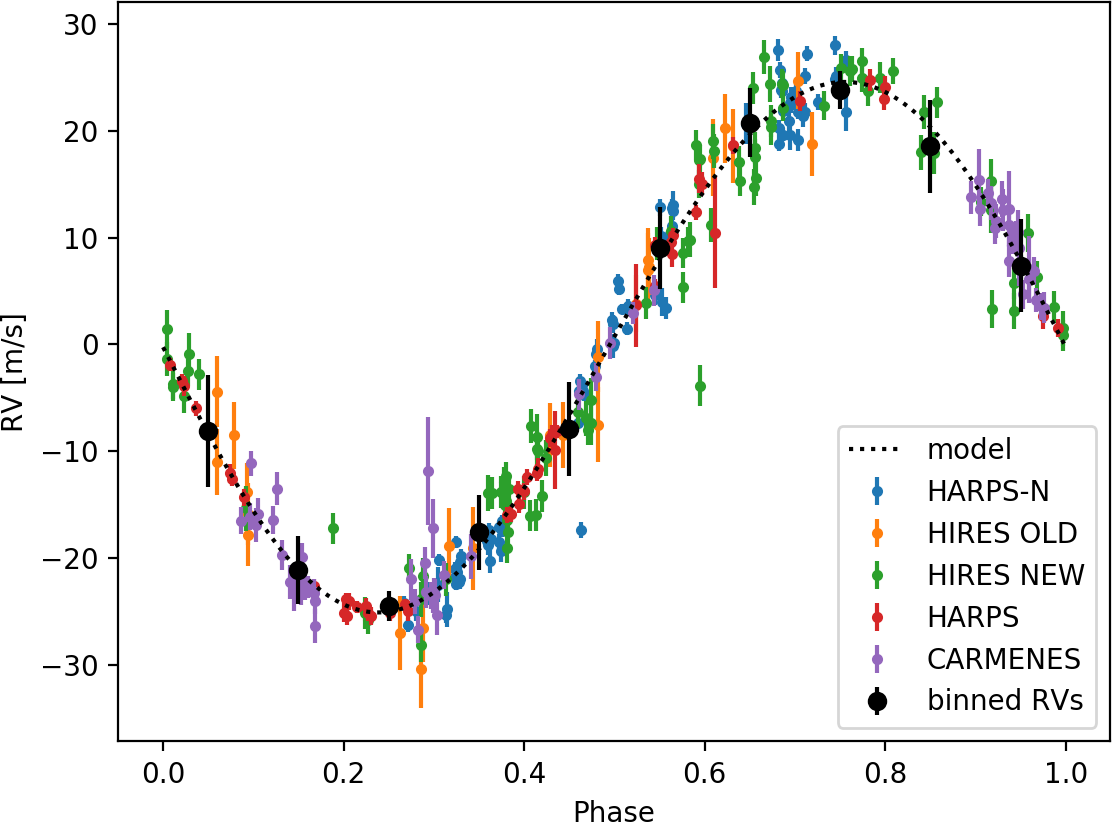}
 \label{fig:gj849_phase1}} \\
   \subfloat[][]
 {\includegraphics[width=.45\textwidth]{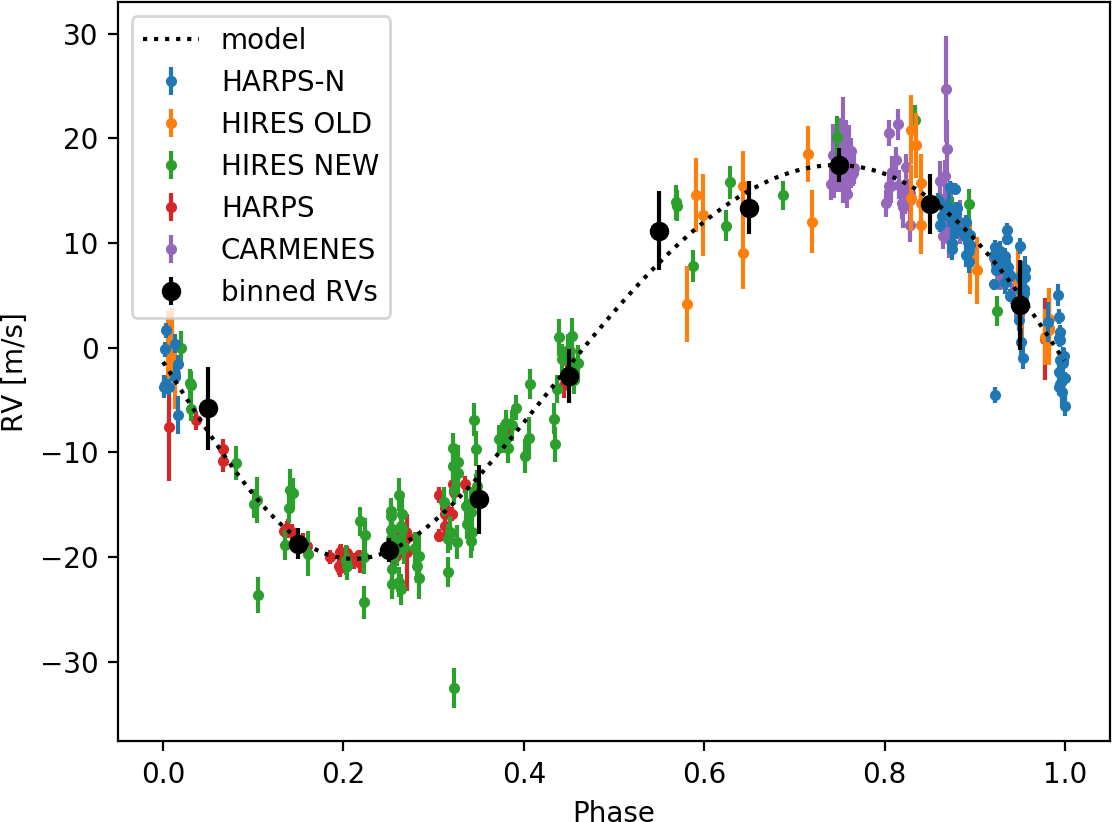}
 \label{fig:gj849_phase2}}
 \caption{GJ 849: \textit{Upper panel:} Phase-folded RV signal of GJ 849\,b, after the subtraction of the stellar correlated signal. \textit{Lower panel:}   Phase-folded RV signal of GJ 849\,c, after the subtraction of the stellar correlated signal.}
 \label{fig:gj849_phase}
\end{figure}

To confirm the stellar nature of the RV short-period signal, we applied the GP regression on the complete RV time series of GJ 849, adopting the quasi-periodic kernel described in Eq. \ref{eq:gp_kernel}. We adopted a large uninformative prior for the rotation period, $\theta$,  $\mathcal{U}$(10,50), in order to  include both the $\simeq20$ d and $\simeq40$ d periodicities identified in the RV residuals and activity indices. This fit results in a clear rotation period of $\theta = 40.45^{+0.19}_{-0.18}$ d, confirm the measurement by \citet{suarezmascarenoetal2015}, and the quasi-periodic model absorbs completely the 20 d signal seen in the previous residuals, confirming that it was related to stellar activity.
The phase folded signals of the 2 planets are shown in Fig. \ref{fig:gj849_phase1} and \ref{fig:gj849_phase2}, the quasi-periodic stellar model is shown in Fig. \ref{fig:gj849_gp_model}, while the complete details of the model and results are discussed in Appendix \ref{app:emcee}.
After correcting the stellar activity via GP regression, we found no significant difference in the orbital parameters of the two planets with respect to the previous fit.  The best-fit orbital parameters of the planetary signals in the final mcmc model are listed in Tab. \ref{tab:all_orbital_parameters}. No additional signal is identified in the RV residuals after the subtraction of the planetary and stellar signals.

\begin{figure}
   \centering
   \subfloat[][]
 {\includegraphics[width=9cm]{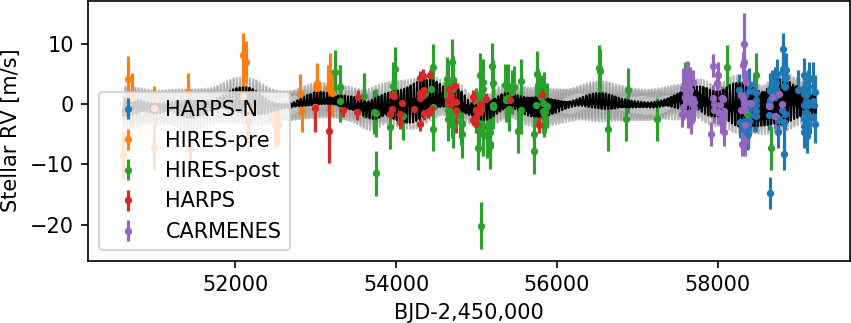}
 \label{fig:gj849_gp_model1}} \\
   \subfloat[][]
 {\includegraphics[width=9cm]{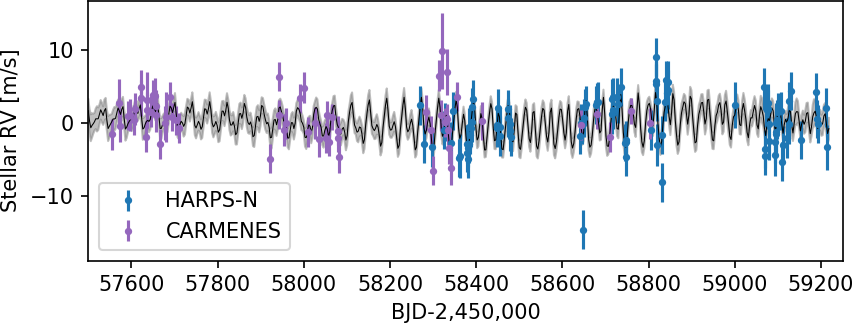}
 \label{fig:gj849_gp_model2}}
   
      \caption{GJ 849: \textit{Upper panel:} best-fit stellar quasi-periodic signal obtained from the GP + 1 planet model (black line) compared to the RV residuals.  \textit{Lower panel: close-up of the CARMENES and HARPS-N time series.}}
         \label{fig:gj849_gp_model}
\end{figure}

\begin{table*}
\caption[]{Best-fit orbital parameters for the planets in the studied systems.}
\label{tab:all_orbital_parameters}
\centering
\begin{tabular}{lcccc}
\hline
\hline
\noalign{\smallskip}
&  GJ 328 & GJ 649 & GJ 849 \\
\noalign{\smallskip}
\hline
\noalign{\smallskip}
$K_b$ $[$m s$^{-1}]$ & $42.6^{+1.8}_{-1.7}$ & $ 9.71^{+0.55}_{-0.53}$ & $24.85^{+0.61}_{-0.63}$  \\
\noalign{\smallskip}
$P_b$ $[$d$]$ & $3771^{+17}_{-17}$ & $600.1^{+1.7}_{-1.7}$ & $1925.31^{+6.5}_{-6.5}$  \\
\noalign{\smallskip}
$T0_b$ $[$BJD$-2450000]$ & $6177^{+45}_{-45}$ & $9045.4^{+10.8}_{-11.7}$ & $3906^{+12}_{-11}$ \\
\noalign{\smallskip}
$M_b \sin i$ $[$M$_\text{J}]$  & $2.51^{+0.23}_{-0.23}$ & $0.258^{+0.023}_{-0.022}$ & $0.893^{+0.094}_{-0.097}$ \\
\noalign{\smallskip}
$a_b$ $[$AU$]$  & $4.11^{+0.16}_{-0.18}$ & $1.112^{+0.035}_{-0.037}$ & $2.32^{+0.11}_{-0.13}$ \\
\noalign{\smallskip}
$e_b$  & $0.227^{+0.015}_{-0.015}$  & $0.083^{+0.068}_{-0.055}$ & $0.029^{+0.019}_{-0.019}$ \\
\noalign{\smallskip}
$\omega_b$ $[$rad$]$ & $-1.33^{+0.11}_{-0.11}$ & $ 0.06^{+0.73}_{-0.71}$ & $ 1.94^{+0.66}_{-0.64}$ \\
\noalign{\smallskip}
\hline
\noalign{\smallskip}
$K_c$ $[$m s$^{-1}]$ & $2.95^{+0.39}_{-0.38}$ & - & $18.81^{+0.81}_{-0.82}$ \\
\noalign{\smallskip}
$P_c$ $[$d$]$ & $241.8^{+1.3}_{-1.7}$ & - & $ 5990^{+110}_{-100}$ \\
\noalign{\smallskip}
$T0_c$ $[$BJD$-2450000]$ & $8478.8^{+5.4}_{-5.3}$ & - & $  3120^{+74}_{-75}$ \\
\noalign{\smallskip}
$M_c \sin i$ $[$M$_\oplus]$  & $21.4^{+ 3.4}_{- 3.2}$ & - & $ 0.99^{+ 0.11}_{- 0.11}$ \\
\noalign{\smallskip}
$a_c$ $[$AU$]$  & $0.657^{+0.026}_{-0.028}$ & - & $ 4.95^{+ 0.25}_{- 0.28}$  \\
\noalign{\smallskip}
$e_c$  & - & - & $0.092^{+0.038}_{-0.036}$  \\
\noalign{\smallskip}
$\omega_c$ $[$rad$]$  & - & -  & $2.49^{+0.44}_{-0.41}$ \\
\noalign{\smallskip}
\hline
\end{tabular}
\end{table*}

\section{Detection limits and planetary occurrence rates}
\label{sec:detection_occurrence}
Although our sample of late-type stars hosting cold Jupiters is limited to four targets, we use it to estimate the planetary occurrence rate in such systems, with a particular focus on the frequency of short-period sub-Neptunes. We do this following the Bayesian approach adopted by \citet{pinamontietal2022}, which takes advantage of the \texttt{emcee} framework described above to compute the detection limits of the RV time series. For a complete description of the statistical approach, see Sect. 3 of \citet{pinamontietal2022}, and references therein.

\begin{figure}
   \centering
   \includegraphics[width=9cm]{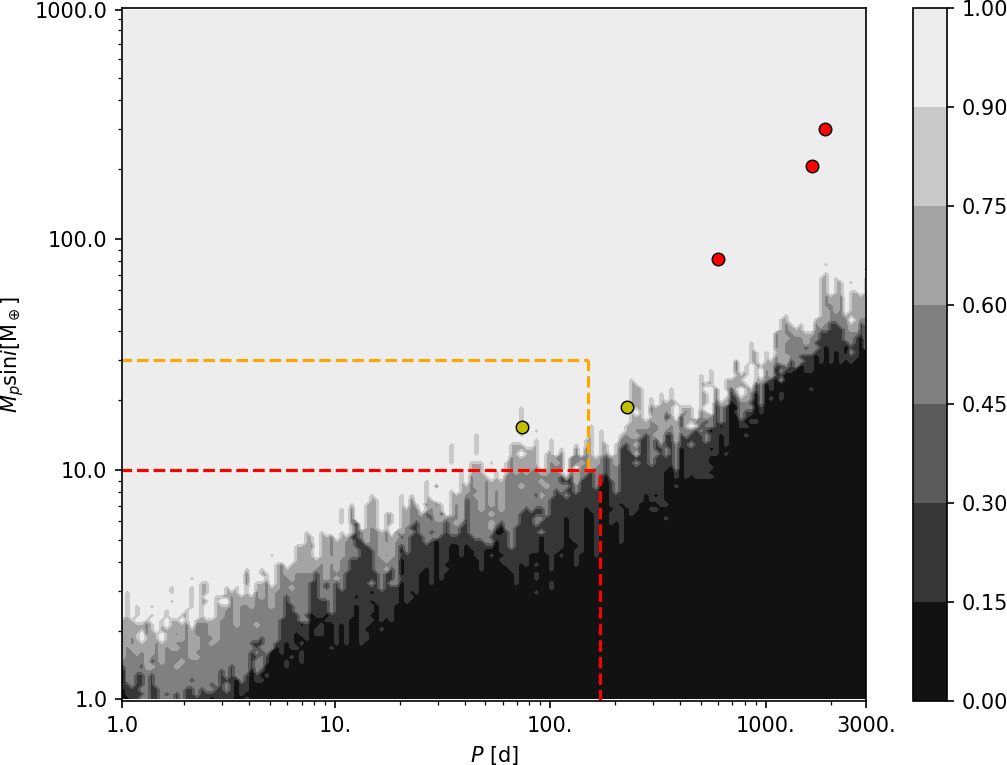}
      \caption{Survey detection map. The grey scale expresses the global detection function, $p$. The red circles mark the position in the parameter space of the known cold Jupiters (see Fig. \ref{fig:systems_lit}), while the yellow circles indicate the short-period planets detected in this study and \citet{barbatoetal2020}. The orange and red dashed contours show the definitions of inner mini-Neptunes by \citet{barbatoetal2018} and super-Earths by \citet{bryanetal2019}.}
         \label{fig:detection_map}
\end{figure}

We computed the detection limits for the three targets analyzed in this work, as well as for BD-11 4672 \citep{barbatoetal2020}. The resulting average detection map is shown in Fig. \ref{fig:detection_map}.

Given the detectability function, $p$, the planetary occurrence rate, $f_\text{occ}$, expressed as the number of planets per star, can be computed from the Poisson distribution:

\begin{equation}
\label{eq:poisson}
 \mathcal{P}(k \mid n, f_\text{occ}) = {(n f_\text{occ})^k e^{-n f_\text{occ}} \over {k!}},
\end{equation}
where $k$ is the number of detected planets, and the expected value is computed as the product between $n$, the number of targets sensitive to planets, and $f_\text{occ}$.

To compute meaningful occurrence rates to test the influence of Cold Jupiters on inner low-mass planet formation, we have to exactly define the intervals in the parameter space corresponding to the definition of inner low-mass planet: unfortunately, the literature on the subject is quite inconsistent, with different definitions being used by different authors, e.g. $M < 10$ M$_\oplus$, $a < 1$ AU \citep{zhuwu2018},  $10$ M$_\oplus < M \sin i < 30$ M$_\oplus$, $P < 150$ d \citep{barbatoetal2018}, $M < 10$ M$_\oplus$, $a < 0.5$ AU \citep{bryanetal2019}, $2$ M$_\oplus < M \sin i < 30$ M$_\oplus$, $0.023 < a < 1$ AU \citep{rosenthaletal2022}. 
We take as reference definitions the ones from \citet{barbatoetal2018} and \citet{bryanetal2019}, as they cover two adjacent ranges of mass and similar period intervals, and we will define them hereafter as mini-Neptunes and super-Earths, respectively. These two intervals are depicted in Fig. \ref{fig:detection_map}.\footnote{\label{foot:ref_mass}Here and in the following Section, we converted all the semi-major axis intervals into orbital period intervals, adopting the mean stellar mass of the sample, $0.565$, as the reference value for the conversion.}
For the mini-Neptunes interval, in which 1 planet was detected around our sample, we obtain an occurrence rate of $\fSNCJ = 0.25^{+0.58}_{-0.07}$, while for the super-Earths bin in which no planet was detected we obtain a 68$\%$ upper limit of $\fSNCJ<0.64$.

To test whether and how the presence of cold Jupiters has a significant effect on the frequency of inner sub-Neptunes, we have to compare the computed occurrence rates with the respective values computed for a sample of field-M dwarfs of similar spectral type, $\fSN$. We adopted the HADES sample of early-M dwarfs as a reference sample \citep{pinamontietal2022}, as it is composed of field M dwarfs of similar masses as our sample of Jupiter-hosting stars. We can thus compute the frequencies of mini-Neptunes and super-Earths around field M dwarfs, which are $\fSN = 0.06^{+0.06}_{-0.02}$ and $\fSN = 0.78^{+0.28}_{-0.16}$, respectively.

The derived estimates of the occurrence rate of inner sub-Neptunes in late-type systems hosting cold Jupiters
show some interesting features: most notably, we can se that for mini-Neptunes ($10$ M$_\oplus < M \sin i < 30$ M$_\oplus$) $\fSNCJ > \fSN$ at a 2$\sigma$ level. Although not strongly significant, this could be an indication that mini-Neptunes are more frequent around late-type stars hosting long-period giant planets, and that a positive correlation exists between the two populations. This is all the more interesting considering that previous studies on solar-type stars observed an opposite behavior: \citet{barbatoetal2018}, studying a sample of 20 solar-type cold-Jupiter hosts, found an upper limit to the frequency of mini-Neptunes of $\fSNCJ < 9.84\%$, significantly lower than the occurrence rates for field solar-type stars, $\fSN = 38.8 \pm 7.1\%$ \citep{mayor2011}.

Considering instead super-Earths ($M \sin i < 10$ M$_\oplus$), we notice that while we detected no planet with such characteristics orbiting our targets, the corrected occurrence rate $\fSNCJ$ could still be compatible with the field frequency $\fSN$, being only different $\simeq 1.2 \sigma$ due to the large uncertainties on the derived upper limit of $\fSNCJ$. Although not significant, this is again a very different behavior than what observed in the literature for solar-type stars, since \citet{bryanetal2019} derived a much higher frequency of sub-Neptunes in systems hosting long-period giant planets, $\fSNCJ >> \fSN$.\footnote{It is worth noticing, however, that more recent studies are not confirming the high $\sim 100\%$ frequencies derived by \citet{bryanetal2019} \citep[e.g.][]{rosenthaletal2022,bonomoetal2023}.}

Considering the other intervals of definition from the literature \citep{zhuwu2018,rosenthaletal2022}, the observed behaviors do not change significantly: computing the occurrence rates for inner super-Earths as defined by \citet{zhuwu2018}, we obtain $\fSNCJ < 0.76$ compatible with the respective $\fSN = 0.94^{+0.34}_{-0.19}$ ($1.2\sigma$). On the other hand, following the definition adopted by \citet{rosenthaletal2022} that includes both high- and low-mass inner companions, we derive $\fSNCJ = 0.78^{+1.02}_{-0.24}$ which is higher but compatible with the corresponding field occurrence rate, $\fSN = 0.61^{+0.19}_{-0.12}$.
It is worth noticing that our results suggest a strong difference in the behavior of late-type and solar-type systems hosting long-period giant planets: this highlights once again the importance of considering the host mass in planetary population studies \citep[e.g.][]{gaidosetal2016,sabottaetal2021,pinamontietal2022}, and points out the risks of drawing conclusions on planetary populations from heterogeneous samples of stars of different masses and spectral types \citep[e.g.][]{bryanetal2019,rosenthaletal2022}.

\section{Discussion and conclusions}
\label{sec:conclusions}

We presented in this work the RV monitoring of 3 early-M dwarfs hosting long-period giant planets, carried out within the GAPS programme. We detected one new low-mass planet around GJ 328, GJ328\,c, with a period of $P_c = 241.8^{+1.3}_{-1.7}$ d and a minimum mass of $M_c \sin i = 21.4^{+ 3.4}_{- 3.2}$ M$_\oplus$: this is a mini-Neptune in the mass range we were looking for, although on a slightly longer orbital period. The other two observed systems, GJ 649 and GJ 849, also showed short-period RV variability, but we confirmed it to be caused by the stellar chromospheric activity, measuring a rotation period of $P_\text{rot} = 24.89^{+0.34}_{-0.35}$ d and $P_\text{rot} = 40.45^{+0.19}_{-0.18}$ d for GJ 649 and GJ 849, respectively. We correctly modelled the activity signals via GP regression, and found no evidence of additional planetary companions in those systems.

Moreover, we updated the orbital parameters of the known cold Jupiters orbiting in these systems. For GJ328\,b we obtained a lower eccentricity than the original values \citep{robertsonetal2013b}, although this difference is not strongly significant ($\simeq 2.7\sigma$). On the other hand, the period and minimum mass are consistent with the literature. For GJ 649\,b we found no evidence of the eccentric orbit, $e=0.3$, reported by \citet{johnsonetal2010} and, after the activity correction, found that the orbital eccentricity is consistent with 0. Finally, for the  GJ 849 system, the GP activity correction confirmed the updated orbital parameters of GJ 849\,b derived by \citet{fengetal2015}, and improved the precision on the orbit of GJ 849\,c.

We then performed a Bayesian analysis of the three systems presented in this work and BD-11 4672 \citep{barbatoetal2020}, in order to estimate the unbiased occurrence rate of low-mass inner planets around late-type stars hosting cold Jupiters. We derived a frequency $\fSNCJ = 0.25^{+0.58}_{-0.07}$ for mini-Neptunes ($10$ M$_\oplus < M \sin i < 30$ M$_\oplus$, $P < 150$ d), which is marginally higher than the frequency for field stars $\fSN$ ($2\sigma$) and might thus be an indication of a positive correlation between the population of cold Jupiters and mini-Neptunes around late-type stars, contrary to what is observed for Sun-like stars. With no planet detected with $ M \sin i < 10$ M$_\oplus$, we measured an upper limit to the frequency of super-Earths $\fSNCJ < 0.64$, which is compatible with the corresponding occurrence rate around field M dwarfs. This is again different than what observed for solar-type stars, where the occurrence rate of super-Earths appears to be boosted by the presence of outer cold Jupiters, although the exact magnitude of this effect is still debated in the literature \citep[e.g.][]{bryanetal2019,rosenthaletal2022}. These results, although limited by the size of our sample, suggest that the formation of hierarchical systems around late-type stars follow a different path than around solar-type stars, as the influence of long-period giant planets appears to be opposite for the two classes of stars. This adds another piece to the puzzle of the dependence of planetary formation on the characteristics of the host stars, in particular on the mass which is known to greatly influence the resulting frequency of planets both in size and orbital separation.

It is worth noticing that our results, although interesting, are limited by the small number of observed objects (4). Although there is an intrinsic physical limitation in analyses such as ours, due to the low frequency of cold Jupiters around M dwarfs \citep[][]{clantongaudi2014}\footnote{Although some preliminary analyses on RV long-term trends suggest that the frequency could be higher \citep[and references therein]{pinamontietal2022}.}, a few additional similar systems are present in the literature and could be added to the sample to improve the statistics\footnote{9 additional M dwarfs hosting RV-detected cold Jupiters are present in the NASA Exoplanet Archive as of 13 January 2023.}.
However, including these systems in the statistical would riquire additional observations, to enhance the sensitivity of the RV time series of such additional targets down to super-Earth masses, and thus is beyond the scope of this current work, which aims to present the data and results of our survey. 

The objective of the presented survey was the intensive monitoring of the 4 observed late-K to M dwarfs hosting known long-period giant planets, in order to allow the detection of low-mass planets in inner orbits. We detected new sub-Neptunes around two of the targets, BD-11 4672\,c \citep{barbatoetal2020} and GJ 328\,c, and excluded the presence of additional short-period companions around the other two targets down to super-Earth masses. Comparing our sensitivity to the previous HARPS Solar-type survey, \citet{barbatoetal2020} achieved detection completeness over their 20-star sample for masses $M\sin i > 30$ M$_\oplus$ for periods below 50 d, and for masses $M\sin i > 50$ M$_\oplus$ for periods below 150 d; considering the same period thresholds, over our sample we are sensitive to masses $M\sin i > 15$ M$_\oplus$ for periods below 50 d, and for masses $M\sin i > 20$ M$_\oplus$ for periods below 150 d. This highlights the importance of high-cadence high-precision RV observations in the study of low-mass short-period planets.

We plan additional observations of other late-type stars, from K to M, some of which are already ongoing with HARPS-N at TNG and FIES at NOT. Moreover, within the GAPS programme a survey of 19 Solar-type host of cold Jupiters was conducted over the recent years, producing the detection of a close-in super-Earth around HD 164922 \citep{benattietal2020}, and additional detections yet to be published. The thorough discussion of the frequencies of small mass planets around cold-Jupiter hosts of different spectral types, from these surveys and future observations, will be the content of future works.

\begin{acknowledgements}
GAPS acknowledges support from INAF through the `Progetti Premiali' funding scheme of the Italian Ministry of Education, University, and Research.
The HARPS-N Project is a collaboration between the Astronomical Observatory of the Geneva University (lead), the CfA in Cambridge, the Universities of St. Andrews and Edinburgh, the Queen’s University of Belfast, and the TNG-INAF Observatory.\\
We acknowledge support from the PRIN-INAF 2019 "Planetary systems at young ages (PLATEA)", ASI-INAF agreement n.2018-16-HH.0, and ASI-INAF n.2021-5-HH.0 ``Partecipazione italiana alla fase B2/C della missione Ariel''. 
MPi acknowledges the financial support from the ASI-INAF Addendum n.2018-24-HH.1-2022 ``Partecipazione italiana al Gaia DPAC  - Operazioni e attivit\`a di analisi dati''.
DB acknowledges the financial support of the National Centre for Competence in Research PlanetS of the Swiss National Science Foundation (SNSF).
L.M. acknowledges support from the ``Fondi di Ricerca Scientifica d’Ateneo 2021'' of the University of Rome ``Tor Vergata''.
DN acknowledges the support from the French Centre National d’Etudes Spatiales (CNES).\\
The HARPS-N Project is a collaboration between the Astronomical Observatory of the Geneva University (lead), the CfA in Cambridge, the Universities of St. Andrews and Edinburgh, the Queen’s University of Belfast, and the TNG-INAF Observatory. This work has made use of data from the European Space Agency (ESA) mission Gaia (\url{https://www.cosmos.esa.int/gaia}), processed by the Gaia Data Processing and Analysis Consortium (DPAC, \url{https://www.cosmos.esa.int/web/gaia/dpac/consortium}). Funding for the DPAC has been provided by national institutions, in particular the institutions participating in the Gaia Multilateral Agreement.\\
This research has made use of NASA’s Astrophysics Data System Bibliographic Services.
This research has made use of the SIMBAD database, operated at CDS, Strasbourg, France.
This research has made use of the NASA Exoplanet Archive, which is operated by the California Institute of Technology, under contract with the National Aeronautics and Space Administration under the Exoplanet Exploration Program.
This research has made use of data retrieved from the SOPHIE archive at Observatoire de Haute-Provence (OHP), available at \url{atlas.obs-hp.fr/sophie}.
This research has made use of observations made with ESO Telescopes at the La Silla Paranal Observatory under programmes IDs 072.C-0488(E) and 183.C-0437(A).\\
MPi and DB also wish to thank A. Baglio, G. Storti, G. Poretti, and M. Massironi for their inspirational work in precision mechanics and advanced technology.
\end{acknowledgements}

  \bibliographystyle{aa} 
  \bibliography{/mnt/DATA/Matteo/Lavoro/articoli/biblio}

\begin{appendix}

\section{\texttt{emcee} priors, posteriors, and best-fit parameters}
\label{app:emcee}

\begin{figure*}
   \centering
   \includegraphics[width=.95\textwidth]{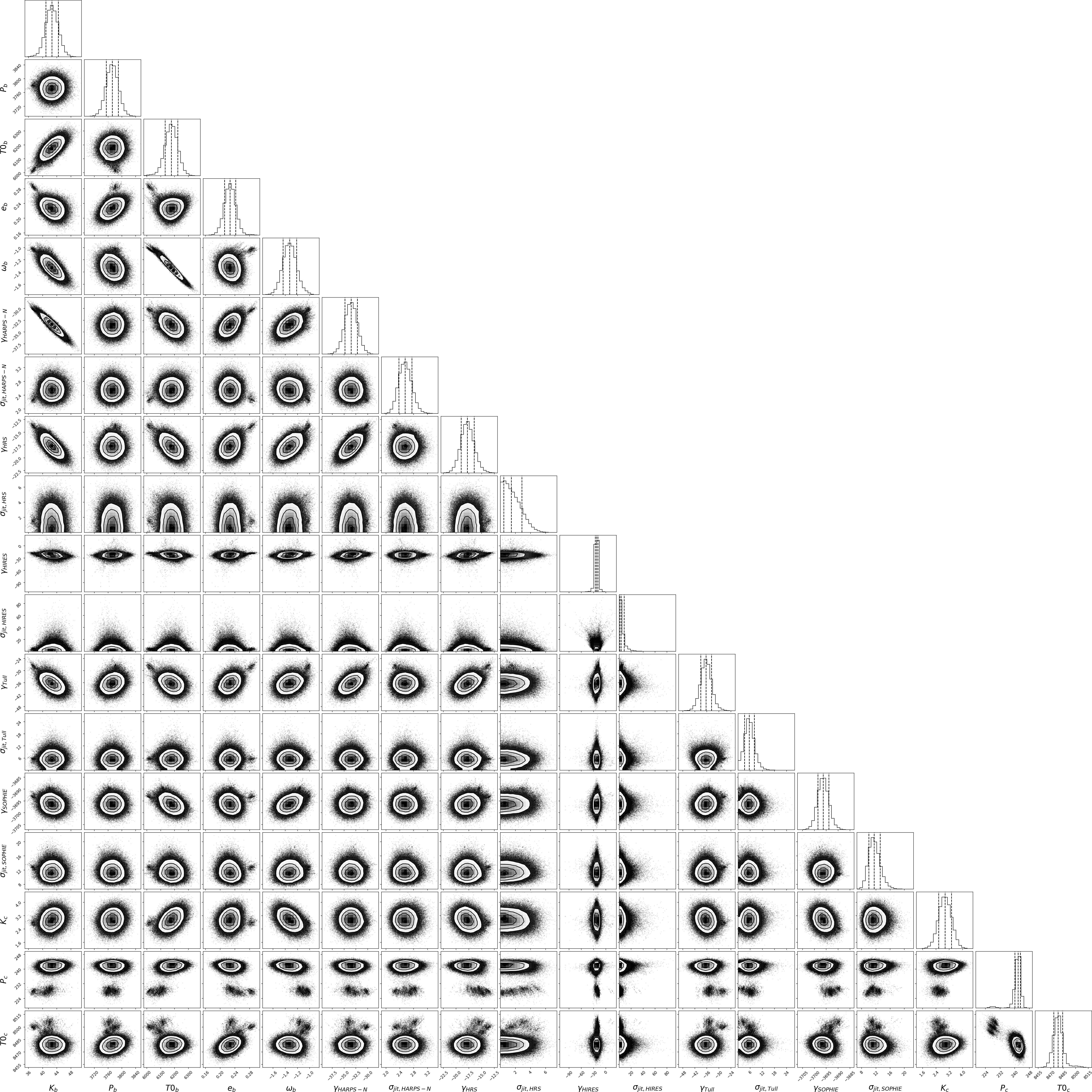}
      \caption{GJ 328: Posterior distributions of the fitted parameters of the 2 Keplerians $(e_c=0)$ model applied to the RV time series. The vertical dashed lines denote the median and the 16th and 84th percentiles. The shown parameters are, in order: $k_b$, $P_b$, $T0_b$, $e_b$, $\omega_b$, $\gamma_\text{HARPS-N}$, $\sigma_\text{jit,HARPS-N}$, $\gamma_\text{HRS}$, $\sigma_\text{jit,HRS}$, $\gamma_\text{HIRES}$, $\sigma_\text{jit,HIRES}$, $\gamma_\text{Tull}$, $\sigma_\text{jit,Tull}$, $\gamma_\text{SOPHIE}$, $\sigma_\text{jit,SOPHIE}$, $k_c$, $P_c$, and $T0_c$.}
         \label{fig:gj328_posteriors}
\end{figure*}

\begin{figure*}
   \centering
   \includegraphics[width=.95\textwidth]{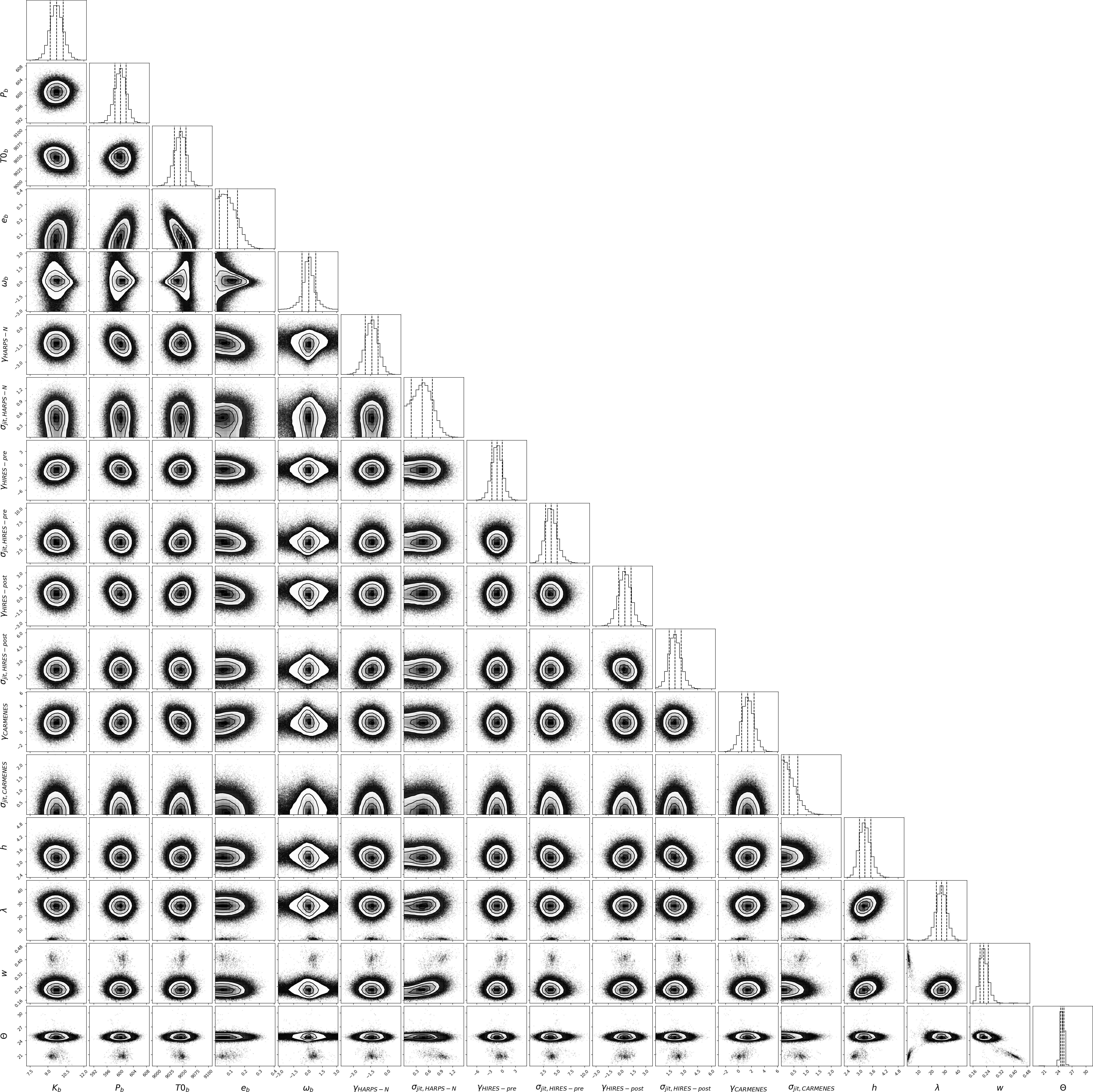}
      \caption{GJ 649: Posterior distributions of the fitted parameters of the Keplerian + GP model applied to the RV time series. The vertical dashed lines denote the median and the 16th and 84th percentiles. The shown parameters are, in order: $K_b$, $P_b$, $T0_b$, $e_b$, $\omega_b$, $\gamma_\text{HARPS-N}$, $\sigma_\text{jit,HARPS-N}$, $\gamma_\text{HIRES-pre}$, $\sigma_\text{jit,HIRES-pre}$, $\gamma_\text{HIRES-post}$, $\sigma_\text{jit,HIRES-post}$, $\gamma_\text{CARMENES}$, $\sigma_\text{jit,CARMENES}$, $h$, $\lambda$, $w$, and $\theta$.}
         \label{fig:gj649_posteriors}
\end{figure*}

\begin{figure*}
   \centering
   \includegraphics[width=.95\textwidth]{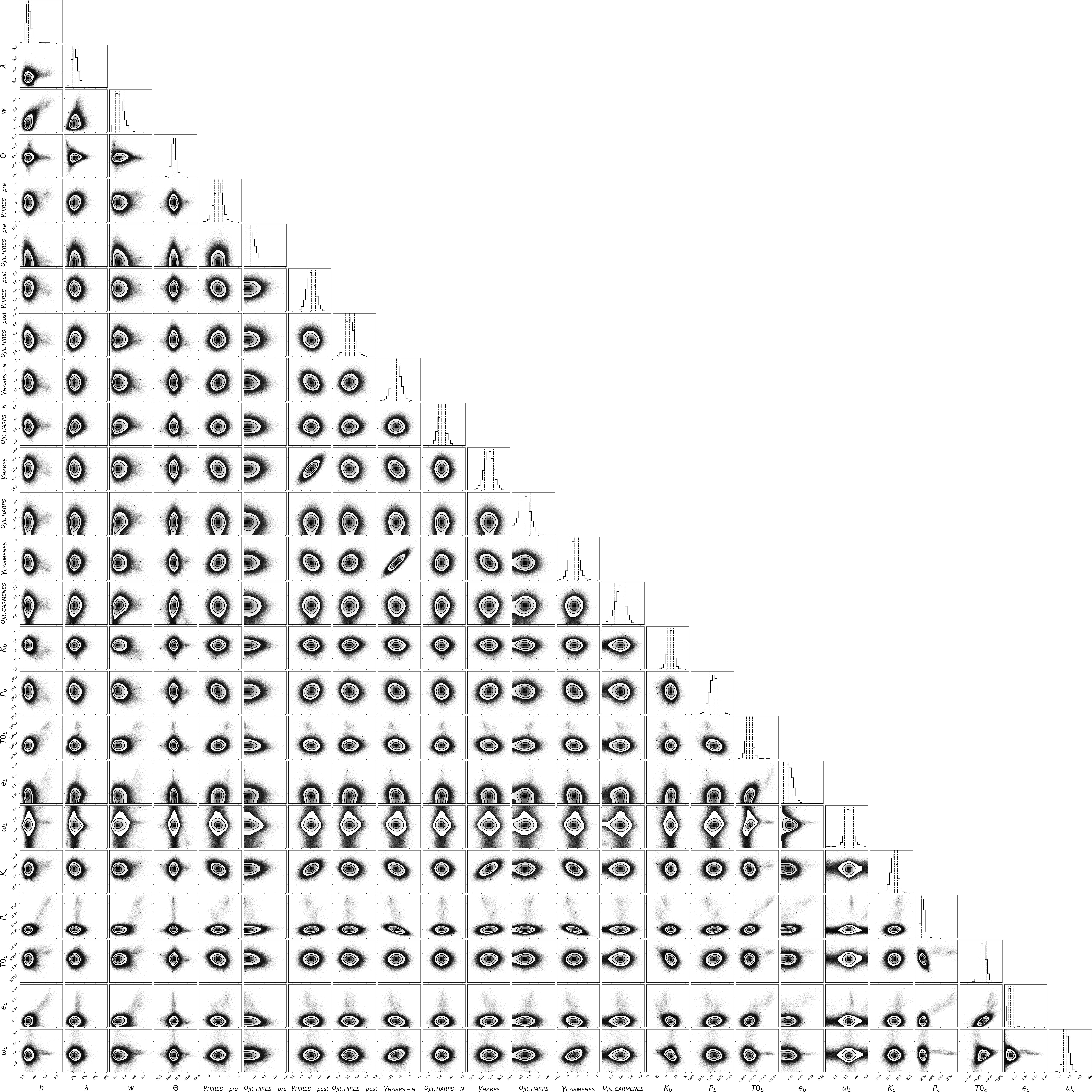}
      \caption{GJ 849: Posterior distributions of the fitted parameters of the 2 Keplerians + GP model applied to the RV time series. The vertical dashed lines denote the median and the 16th and 84th percentiles.  The shown parameters are, in order:$h$, $\lambda$, $w$, $\theta$, $\gamma_\text{HIRES-pre}$, $\sigma_\text{jit,HIRES-pre}$, $\gamma_\text{HIRES-post}$, $\sigma_\text{jit,HIRES-post}$, $\gamma_\text{HARPS-N}$, $\sigma_\text{jit,HARPS-N}$, $\gamma_\text{HARPS}$, $\sigma_\text{jit,HARPS}$, $\gamma_\text{CARMENES}$, $\sigma_\text{jit,CARMENES}$, $K_b$, $P_b$, $T0_b$, $e_b$, $\omega_b$, $K_c$, $P_c$, $T0_c$, $e_c$, and $\omega_c$.}
         \label{fig:gj849_posteriors}
\end{figure*}

In this Section, we report the priors and best-fit parameters for all the tested models on the analyzedsystems, and all the posteriors distributions. Tab. \ref{tab:gj328_mcmc_prior_param} lists the information for the three models tested on GJ 328 RV time series, and Fig. \ref{fig:gj328_posteriors} shows the posterior distributions of the fitted parameters in the final 2 Keplerians $(e_c=0)$ model. Tab. \ref{tab:gj649_mcmc_prior_param} lists the details of the models tested on GJ 649 RV time series, and Fig. \ref{fig:gj649_posteriors} shows the posterior distributions of the final Keplerian + GP model. Tab. \ref{tab:gj849_mcmc_prior_param} lists the priors and best-fit parameters of the models applied to GJ 849 RV data, and the posterior distributions of the final 2-Keplerian + GP model are shown in Fig. \ref{fig:gj849_posteriors}.

\begin{table*}
\caption[]{GJ 328: Priors and best-fit parameters for the tested MCMC models. The chosen final model is highlighted in bold.}
\label{tab:gj328_mcmc_prior_param}
\centering
\begin{tabular}{lcccc}
\hline
\hline
\noalign{\smallskip}
 & Priors & \multicolumn{3}{c}{Best-fit parameters}\\ 
&  & 1 Keplerian & 2 Keplerians & \textbf{2 Keplerians $(e_c = 0)$} \\
\noalign{\smallskip}
\hline
$K_b$ $[$m s$^{-1}]$ & $\mathcal{U}$(0,200) & $40.8^{+1.7}_{-1.6}$ & $42.9^{+1.7}_{-1.7}$ & $42.6^{+1.8}_{-1.7}$ \\ 
\noalign{\smallskip}  
$P_b$ $[$d$]$ & $\mathcal{U}$(3000,4000) & $3763^{+17}_{-17}$ & $3772^{+18}_{-17}$ & $3771^{+17}_{-17}$ \\ 
\noalign{\smallskip}  
$T0_b$ $[$BJD$-2450000]$ & $\mathcal{U}$(5600,6900) & $7191^{+44}_{-41}$ & $6177^{+44}_{-45}$ & $6177^{+45}_{-45}$ \\
\noalign{\smallskip}  
$\sqrt{e_b} \cos{\omega_b}$ & $\mathcal{U}$(-1,1) & $0.230^{+0.043}_{-0.047}$ & $0.112^{+0.051}_{-0.050}$ & $0.114^{+0.051}_{-0.050}$ \\
\noalign{\smallskip}
$\sqrt{e_b} \sin{\omega_b}$ & $\mathcal{U}$(-1,1) & $-0.412^{+0.028}_{-0.026}$ & $-0.458^{+0.024}_{-0.021}$ & $-0.460^{+0.023}_{-0.020}$ \\
\noalign{\smallskip}
\hline
\noalign{\smallskip}  
$M_b \sin i$ $[$M$_\text{J}]$  &  & $2.40^{+0.22}_{-0.22}$ & $2.53^{+0.23}_{-0.23}$  & $2.51^{+0.23}_{-0.23}$ \\ 
\noalign{\smallskip}  
$a_b$ $[$AU$]$  &  & $4.10^{+0.16}_{-0.18}$ & $4.11^{+0.16}_{-0.18}$  & $4.11^{+0.16}_{-0.18}$ \\ 
\noalign{\smallskip}  
$e_b$  &  & $0.224^{+0.015}_{-0.015}$  & $0.225^{+0.015}_{-0.016}$ & $0.227^{+0.015}_{-0.015}$ \\ 
\noalign{\smallskip}  
$\omega_b$ $[$rad$]$ &  & $-1.06^{+0.10}_{-0.11}$ &$-1.33^{+0.12}_{-0.11}$ & $-1.33^{+0.11}_{-0.11}$ \\ 
\noalign{\smallskip}  
\hline  
\noalign{\smallskip}  
$K_c$ $[$m s$^{-1}]$ & $\mathcal{U}$(0,50) & - & $3.18^{+0.51}_{-0.46}$ & $2.95^{+0.39}_{-0.38}$ \\ 
\noalign{\smallskip}  
$P_c$ $[$d$]$ & $\mathcal{U}$(210,250) & - & $242.5^{+1.3}_{-1.7}$ & $241.8^{+1.3}_{-1.7}$ \\ 
\noalign{\smallskip}  
$T0_c$ $[$BJD$-2450000]$ & $\mathcal{U}$(8400,8600) & - & $8485^{+10}_{-11}$ & $8478.8^{+5.4}_{-5.3}$ \\ 
\noalign{\smallskip}  
$\sqrt{e_c} \cos{\omega_c}$ & $\mathcal{U}$(-1,1) & - & $-0.17^{+0.23}_{-0.19}$ & - \\ 
\noalign{\smallskip}  
$\sqrt{e_c} \sin{\omega_c}$ & $\mathcal{U}$(-1,1) & - & $-0.33^{+0.43}_{-0.20}$ & - \\ 
\noalign{\smallskip}  
\hline  
\noalign{\smallskip}  
$M_c \sin i$ $[$M$_\oplus]$  &  & - & $24.0^{+ 4.9}_{- 4.1}$ & $21.4^{+ 3.4}_{- 3.2}$ \\ 
\noalign{\smallskip}  
$a_c$ $[$AU$]$  &  & - & $0.659^{+0.026}_{-0.028}$ & $0.657^{+0.026}_{-0.028}$ \\ 
\noalign{\smallskip}  
$e_c$  &  & - & $0.22^{+0.15}_{-0.13}$ & - \\ 
\noalign{\smallskip}  
 &  &  & $(< 0.28)$ & - \\ 
\noalign{\smallskip}  
$\omega_c$ $[$rad$]$  & & - & $-1.76^{+3.51}_{-0.63}$ & - \\ 
\noalign{\smallskip}  
\hline  
\noalign{\smallskip}  
$\gamma_\text{HARPS-N}$ $[$m s$^{-1}]$ & $\mathcal{U}$(-200,200) & $-33.3^{+1.3}_{-1.3}$ & $-33.8^{+1.3}_{-1.3}$ & $-33.5^{+1.3}_{-1.3}$ \\ 
\noalign{\smallskip}  
$\sigma_\text{jit,HARPS-N}$ $[$m s$^{-1}]$ & $\mathcal{U}$(0,150) & $3.12^{+0.23}_{-0.21}$ & $2.54^{+0.20}_{-0.18}$ & $2.55^{+0.19}_{-0.18}$ \\ 
\noalign{\smallskip}  
$\gamma_\text{HRS}$ $[$m s$^{-1}]$ & $\mathcal{U}$(-200,200) & $-16.2^{+1.2}_{-1.2}$ & $-17.9^{+1.3}_{-1.2}$ & $-17.8^{+1.3}_{-1.3}$ \\ 
\noalign{\smallskip}  
$\sigma_\text{jit,HRS}$ $[$m s$^{-1}]$ & $\mathcal{U}$(0,150) & $1.9^{+1.5}_{-1.3}$ & $1.6^{+1.4}_{-1.1}$ & $1.5^{+1.4}_{-1.0}$ \\ 
\noalign{\smallskip}  
$\gamma_\text{HIRES}$ $[$m s$^{-1}]$ & $\mathcal{U}$(-200,200) & $-22.2^{+3.2}_{-3.6}$ & $-22.6^{+3.0}_{-3.1}$ & $-22.5^{+3.0}_{-3.2}$ \\ 
\noalign{\smallskip}  
$\sigma_\text{jit,HIRES}$ $[$m s$^{-1}]$ & $\mathcal{U}$(0,150) & $4.1^{+6.0}_{-2.8}$ & $3.1^{+4.5}_{-2.2}$ & $3.2^{+4.9}_{-2.3}$ \\ 
\noalign{\smallskip}  
$\gamma_\text{Tull}$ $[$m s$^{-1}]$ & $\mathcal{U}$(-200,200) & $-34.6^{+2.6}_{-2.6}$ & $-36.5^{+2.5}_{-2.5}$ & $-36.4^{+2.7}_{-2.7}$ \\ 
\noalign{\smallskip}  
$\sigma_\text{jit,Tull}$ $[$m s$^{-1}]$ & $\mathcal{U}$(0,150) & $5.3^{+2.6}_{-2.4}$ & $5.3^{+2.5}_{-2.3}$ & $5.5^{+2.5}_{-2.3}$ \\ 
\noalign{\smallskip}  
$\gamma_\text{SOPHIE}$ $[$m s$^{-1}]$ & $\mathcal{U}$(-3900,3500) & $-3693.4^{+2.2}_{-2.2}$ & $-3696.2^{+2.4}_{-2.3}$ & $-3696.3^{+2.4}_{-2.3}$ \\ 
\noalign{\smallskip}  
$\sigma_\text{jit,SOPHIE}$ $[$m s$^{-1}]$ & $\mathcal{U}$(0,150) & $11.7^{+1.7}_{-1.4}$ & $11.5^{+1.7}_{-1.5}$ & $11.5^{+1.7}_{-1.4}$ \\ 
\noalign{\smallskip}  
\hline  
\noalign{\smallskip}  
BIC & & $1530$ & $1496$ & $1488$ \\ 
\noalign{\smallskip}
\hline
\end{tabular}
\end{table*}

\begin{table*}
\caption[]{GJ 649: Priors and best-fit parameters for the tested MCMC models. The chosen final model is highlighted in bold.}
\label{tab:gj649_mcmc_prior_param}
\centering
\begin{tabular}{lccc}
\hline
\hline
\noalign{\smallskip}
 & Priors & \multicolumn{2}{c}{Best-fit parameters}\\ 
&  & 1 Keplerian & \textbf{GP + 1 Keplerian} \\
\noalign{\smallskip}
\hline
\noalign{\smallskip}
$K_b$ $[$m s$^{-1}]$ & $\mathcal{U}$(0,50) & $ 9.78^{+0.34}_{-0.33}$ & $ 9.71^{+0.55}_{-0.53}$ \\ 
\noalign{\smallskip}  
$P_b$ $[$d$]$ & $\mathcal{U}$(300,1000) & $599.9^{+1.5}_{-1.6}$ & $600.1^{+1.7}_{-1.7}$ \\ 
\noalign{\smallskip}  
$T0_b$ $[$BJD$-2450000]$ & $\mathcal{U}$(8700,9200) & $9040.0^{+6.8}_{-6.9}$ & $9045.4^{+10.8}_{-11.7}$ \\ 
\noalign{\smallskip}  
$\sqrt{e_b} \cos{\omega_b}$ & $\mathcal{U}$(-1,1) & $ 0.28^{+ 0.08}_{- 0.12}$ & $ 0.24^{+ 0.13}_{- 0.20}$ \\ 
\noalign{\smallskip}  
$\sqrt{e_b} \sin{\omega_b}$ & $\mathcal{U}$(-1,1) & $  0.16^{+ 0.08}_{- 0.09}$ & $  0.01^{+ 0.14}_{- 0.15}$ \\ 
\noalign{\smallskip}  
\hline  
\noalign{\smallskip}  
$M_b \sin i$ $[$M$_\text{J}]$  &  & $0.261^{+0.020}_{-0.019}$ & $0.258^{+0.023}_{-0.022}$ \\ 
\noalign{\smallskip}  
$a_b$ $[$AU$]$  &  & $1.112^{+0.035}_{-0.037}$ & $1.112^{+0.035}_{-0.037}$ \\ 
\noalign{\smallskip}  
$e_b$  &  & $0.114^{+0.041}_{-0.041}$  & $0.083^{+0.068}_{-0.055}$ \\ 
\noalign{\smallskip}  
$\omega_b$ $[$rad$]$ &  & $ 0.52^{+0.43}_{-0.32}$ & $ 0.06^{+0.73}_{-0.71}$ \\
\noalign{\smallskip}  
\hline  
$h$ $[$m s$^{-1}]$ & $\mathcal{U}$(0,10) & - & $3.23^{+0.28}_{-0.25}$  \\ 
\noalign{\smallskip}  
$\lambda$ $[$d$]$ & $\log \mathcal{U}$(1,500) & - & $27.4^{+ 3.9}_{- 3.9}$  \\ 
\noalign{\smallskip}  
$w$ & $\mathcal{U}$(0,1) & - & $0.223^{+0.027}_{-0.022}$  \\ 
\noalign{\smallskip}  
$\theta$ $[$d$]$ & $\mathcal{U}$(10,50) & - & $24.89^{+0.34}_{-0.35}$  \\ 
\noalign{\smallskip}  
\hline  
\noalign{\smallskip}  
$\gamma_\text{HIRES-pre}$ $[$m s$^{-1}]$ & $\mathcal{U}$(-50,50) & $ -1.3^{+1.1}_{-1.1}$ & $ -1.3^{+1.2}_{-1.2}$ \\ 
\noalign{\smallskip}  
$\sigma_\text{jit,HIRES-pre}$ $[$m s$^{-1}]$ & $\mathcal{U}$(0,30) & $4.96^{+0.97}_{-0.77}$ & $3.88^{+1.12}_{-0.98}$ \\ 
\noalign{\smallskip}  
$\gamma_\text{HIRES-post}$ $[$m s$^{-1}]$ & $\mathcal{U}$(-50,50) & $ 0.15^{+0.72}_{-0.71}$ & $ 0.47^{+0.75}_{-0.75}$ \\ 
\noalign{\smallskip}  
$\sigma_\text{jit,HIRES-post}$ $[$m s$^{-1}]$ & $\mathcal{U}$(0,30) & $4.23^{+0.55}_{-0.46}$ & $2.06^{+0.66}_{-0.61}$ \\ 
\noalign{\smallskip}  
$\gamma_\text{HARPS-N}$ $[$m s$^{-1}]$ & $\mathcal{U}$(-50,50) & $-1.25^{+0.33}_{-0.33}$ & $-1.40^{+0.56}_{-0.56}$ \\ 
\noalign{\smallskip}  
$\sigma_\text{jit,HARPS-N}$ $[$m s$^{-1}]$ & $\mathcal{U}$(0,30) & $3.11^{+0.21}_{-0.19}$ & $0.45^{+0.25}_{-0.27}$ \\ 
\noalign{\smallskip}  
$\gamma_\text{CARMENES}$ $[$m s$^{-1}]$ & $\mathcal{U}$(-50,50) & $ 1.42^{+0.54}_{-0.54}$ & $ 1.40^{+0.96}_{-0.95}$ \\ 
\noalign{\smallskip}  
$\sigma_\text{jit,CARMENES}$ $[$m s$^{-1}]$ & $\mathcal{U}$(0,30) & $2.66^{+0.39}_{-0.35}$ & $0.31^{+0.34}_{-0.22}$ \\ 
\noalign{\smallskip}  
\hline  
\noalign{\smallskip}  
BIC & & $1497$ & $1370$ \\ 
\noalign{\smallskip}
\hline
\end{tabular}
\end{table*}

\begin{table*}
\caption[]{GJ 849: Priors and best-fit parameters for the tested MCMC models. The chosen final model is highlighted in bold.}
\label{tab:gj849_mcmc_prior_param}
\centering
\begin{tabular}{lccc}
\hline
\hline
\noalign{\smallskip}
 & Priors & \multicolumn{2}{c}{Best-fit parameters}\\ 
&  & 2 Keplerians & \textbf{GP + 2 Keplerians} \\
\noalign{\smallskip}
\hline
\noalign{\smallskip}
$K_b$ $[$m s$^{-1}]$ & $\mathcal{U}$(0,40) & $24.82^{+0.40}_{-0.50}$ & $24.85^{+0.61}_{-0.63}$ \\ 
\noalign{\smallskip}  
$P_b$ $[$d$]$ & $\mathcal{U}$(1800,1980) & $1925.73^{+5.8}_{-5.6}$ & $1925.31^{+6.5}_{-6.5}$ \\ 
\noalign{\smallskip}  
$T0_b$ $[$BJD$-2450000]$ & $\mathcal{U}$(3800,5300) & $3904^{+12}_{-10}$ & $3906^{+12}_{-11}$ \\
\noalign{\smallskip}  
$\sqrt{e_b} \cos{\omega_b}$ & $\mathcal{U}$(-1,1) & $-0.074^{+0.060}_{-0.061}$ & $-0.054^{+0.086}_{-0.075}$ \\ 
\noalign{\smallskip}  
$\sqrt{e_b} \sin{\omega_b}$ & $\mathcal{U}$(-1,1) & $ 0.186^{+0.038}_{-0.047}$ & $ 0.139^{+0.059}_{-0.111}$ \\ 
\noalign{\smallskip}  
\hline  
\noalign{\smallskip}  
$M_b \sin i$ $[$M$_\text{J}]$  &  & $1.013^{+0.079}_{-0.093}$ & $0.893^{+0.094}_{-0.097}$ \\ 
\noalign{\smallskip}  
$a_b$ $[$AU$]$  &  & $2.48^{+0.09}_{-0.09}$ & $2.32^{+0.11}_{-0.13}$ \\ 
\noalign{\smallskip}  
$e_b$  &  & $0.043^{+0.016}_{-0.014}$  & $0.029^{+0.019}_{-0.019}$ \\ 
\noalign{\smallskip}  
$\omega_b$ $[$rad$]$ &  & $ 1.94^{+0.34}_{-0.30}$ & $ 1.94^{+0.66}_{-0.64}$ \\
\noalign{\smallskip}  
\hline  
\noalign{\smallskip}  
$K_c$ $[$m s$^{-1}]$ & $\mathcal{U}$(0,30)  & $18.34^{+0.84}_{-0.77}$ & $18.81^{+0.81}_{-0.82}$ \\ 
\noalign{\smallskip}  
$P_c$ $[$d$]$ & $\mathcal{U}$(4000,8000)  & $ 5970^{+150}_{-110}$ & $ 5990^{+110}_{-100}$ \\ 
\noalign{\smallskip}  
$T0_c$ $[$BJD$-2450000]$ & $\mathcal{U}$(2500,5500)  & $  3119^{+70}_{-63}$ & $  3120^{+74}_{-75}$ \\
\noalign{\smallskip}  
$\sqrt{e_c} \cos{\omega_c}$ & $\mathcal{U}$(-1,1)  & $-0.230^{+0.074}_{-0.062}$  & $-0.23^{+0.10}_{-0.07}$  \\ 
\noalign{\smallskip}  
$\sqrt{e_c} \sin{\omega_c}$ & $\mathcal{U}$(-1,1)  & $0.10^{+0.12}_{-0.13}$ & $0.18^{+0.09}_{-0.12}$  \\ 
\noalign{\smallskip}  
\hline  
\noalign{\smallskip}  
$M_c \sin i$ $[$M$_\oplus]$  &   & $ 1.11^{+ 0.12}_{- 0.10}$ & $ 0.99^{+ 0.11}_{- 0.11}$ \\ 
\noalign{\smallskip}  
$a_c$ $[$AU$]$  &   & $ 5.29^{+ 0.26}_{- 0.22}$ & $ 4.95^{+ 0.25}_{- 0.28}$ \\ 
\noalign{\smallskip}  
$e_c$  &   & $0.071^{+0.033}_{-0.027}$ & $0.092^{+0.038}_{-0.036}$  \\ 
\noalign{\smallskip}  
$\omega_c$ $[$rad$]$  &  & $2.53^{+0.36}_{-5.24}$  & $2.49^{+0.44}_{-0.41}$  \\ 
\noalign{\smallskip}  
\hline  
$h$ $[$m s$^{-1}]$ & $\mathcal{U}$(0,10) & - & $2.24^{+0.36}_{-0.29}$  \\ 
\noalign{\smallskip}  
$\lambda$ $[$d$]$ & $\log \mathcal{U}$(1,500) & - & $ 221^{+  59}_{-  49}$  \\ 
\noalign{\smallskip}  
$w$ & $\mathcal{U}$(0,1) & - & $0.28^{+0.10}_{-0.08}$  \\ 
\noalign{\smallskip}  
$\theta$ $[$d$]$ & $\mathcal{U}$(10,50) & - & $40.45^{+0.19}_{-0.18}$  \\ 
\noalign{\smallskip}  
\hline  
\noalign{\smallskip}  
$\gamma_\text{HIRES-pre}$ $[$m s$^{-1}]$ & $\mathcal{U}$(-100,100) & $  9.3^{+1.3}_{-1.2}$ & $  8.8^{+1.2}_{-1.2}$ \\ 
\noalign{\smallskip}  
$\sigma_\text{jit,HIRES-pre}$ $[$m s$^{-1}]$ & $\mathcal{U}$(0,30) & $ 3.3^{+ 1.4}_{- 1.2}$ & $ 1.5^{+ 1.4}_{- 1.0}$ \\ 
\noalign{\smallskip}  
$\gamma_\text{HIRES-post}$ $[$m s$^{-1}]$ & $\mathcal{U}$(-100,100) & $  5.90^{+0.88}_{-0.71}$ & $  6.15^{+0.72}_{-0.72}$ \\ 
\noalign{\smallskip}  
$\sigma_\text{jit,HIRES-post}$ $[$m s$^{-1}]$ & $\mathcal{U}$(0,20) & $4.02^{+0.53}_{-0.36}$ & $3.35^{+0.40}_{-0.35}$ \\ 
\noalign{\smallskip}  
$\gamma_\text{HARPS-N}$ $[$m s$^{-1}]$ & $\mathcal{U}$(-100,100) & $ -9.1^{+1.6}_{-1.5}$ & $-10.1^{+1.4}_{-1.3}$ \\ 
\noalign{\smallskip}  
$\sigma_\text{jit,HARPS-N}$ $[$m s$^{-1}]$ & $\mathcal{U}$(0,15) & $3.47^{+0.39}_{-0.28}$ & $2.53^{+0.27}_{-0.25}$ \\ 
\noalign{\smallskip}  
$\gamma_\text{HARPS}$ $[$m s$^{-1}]$ & $\mathcal{U}$(-100,100) & $ 26.37^{+0.89}_{-0.80}$ & $ 26.70^{+0.73}_{-0.74}$ \\ 
\noalign{\smallskip}  
$\sigma_\text{jit,HARPS}$ $[$m s$^{-1}]$ & $\mathcal{U}$(0,15) & $1.90^{+0.32}_{-0.25}$ & $0.72^{+0.31}_{-0.33}$ \\ 
\noalign{\smallskip}  
$\gamma_\text{CARMENES}$ $[$m s$^{-1}]$ & $\mathcal{U}$(-100,100) & $ -5.5^{+1.8}_{-1.4}$ & $ -7.0^{+1.3}_{-1.3}$ \\ 
\noalign{\smallskip}  
$\sigma_\text{jit,CARMENES}$ $[$m s$^{-1}]$ & $\mathcal{U}$(0,15) & $2.45^{+0.48}_{-0.33}$ & $1.52^{+0.40}_{-0.44}$ \\ 
\noalign{\smallskip}  
\hline  
\noalign{\smallskip}  
BIC & & $1807$ & $1776$ \\ 
\noalign{\smallskip}
\hline
\end{tabular}
\end{table*}

\end{appendix}

\end{document}